\documentclass[twocolumn,showpacs,preprintnumbers,amsmath,amssymb]{revtex4}

\usepackage{graphicx}
\usepackage{amssymb}
\usepackage{longtable}

\usepackage{epstopdf}
\usepackage{float}      
\newcommand{\ignore}[1]{}
\newcommand{\be}{\begin{equation}} \newcommand{\ee}{\end{equation}}
\newcommand{\ba}{\begin{eqnarray}} \newcommand{\ea}{\end{eqnarray}}
\newcommand{\nn}{\nonumber} \renewcommand{\bf}{bf}
\newcommand{\ra}{\rightarrow}

\renewcommand{\a}{\alpha}

\def\slashb#1{\setbox0=\hbox{$#1$}#1\hskip-\wd0\dimen0=5pt\advance
        \dimen0 by-\ht0\advance\dimen0 by\dp0\lower0.5\dimen0\hbox
          to\wd0{\hss\sl/\/\hss}}

\input{epsf}

\begin{document}
 
\title{One Model Explains  \\ DAMA/LIBRA, CoGENT, \\ CDMS, and XENON} 

\author{John P. Ralston}

  \affiliation{Department of Physics \& Astronomy, \\ The University of Kansas,
  Lawrence, KS 66045}
  
\begin{abstract} Many experiments seek dark matter by detecting relatively low energy nuclear recoils.  Yet since events from ordinary physics with energies in the 1-100 KeV range are commonplace, all claims of signals or their absence hinge on exhaustive calibrations and background rejection. 
We document many curious and consistent discrepancies between the backgrounds which neutrons can produce versus the picture of neutrons and claims of neutron calibration found in dark matter literature. Much of the actual physics of neutrons is either under-recognized or under-reported, opening up new interpretations of current data.  All signals seen so far, including those presented tentatively such as CoGENT, or the bold claims and time dependence of DAMA/LIBRA, appear to be consistent with neutron-induced backgrounds. At the same time it is the burden of proof of experimental groups to support their claims no possible background could matter, not ours.  The existing hypotheses about backgrounds stated by experiments, accepted at face value and as published, make possible a variety of 
neutron-induced events to be registered as dark matter signals.   

\end{abstract}

\maketitle

\section*{The Background Is the Hypothesis } 

Recently three experiments seeking direct detection of particle dark matter have 
reported potential signals. The data of DAMA/LIBRA\cite{dama2010}, CoGENT
\cite{cogent2010}, and CDMS\cite{CDMS2009} are sufficiently different that no 
theoretical model to explain all experiments has come forth. Even more recently, the 
XENON experiment \cite{xenon2010} seems to undercut the others by finding no signal.  
The range of data reported does not seem to be consistent. Here we discuss a model 
for all the data available, with the remarkable feature that everything observed is 
explained without invoking any new physics. 

Our model is {\it neutron backgrounds}.  Neutrons are often cited in experimental 
reports. But there are odd discrepancies between what neutrons do in Nature and how they are 
treated in the literature. Rumors that neutrons are under control have been greatly exaggerated. Neutrons 
may explain all ``signals '' found so far. The news has some positive elements: Flaws in 
assessing 
neutron reactions may also be causing rejection of actual dark matter events. 

Our ``hypothesis'' for the data comes from numerous questions that have not been 
raised or answered in the public record.  It is important that debate over the 
background phenomenology of dark matter searches cannot be restricted to 
collaboration insiders. In the usual case ``discovery'' experiments find new phenomena, 
permitting backgrounds to be hashed out by the insiders of large collaborations.  Yet 
backgrounds are everyone's business when backgrounds become the sole criteria of 
finding new physics.

The premises of dark matter searches are unique.  Experiments define ``signals'' to be 
``events not accounted for by known backgrounds.''  {\it This matters.} The events 
themselves are very ordinary.  Unlike other tests of fundamental physics, finding an 
event does not violate a symmetry. Events can come from myriad causes. No 
experiment can rule out the hypothesis of particle dark matter: it is constructed with 
total freedom of mass, flux, and interactions so that not detecting anything has no effect.  The 
null hypothesis of ``no signal'' of dark matter is also poorly defined.  It is tantamount 
to a model of the background.  

Very conservative early approaches did not need a sharp background model. They could 
treat all energy pulses as potential backgrounds, count them, and set upper limits on dark 
matter interactions. To do better than that, more definition has been given to the null 
hypotheses, via more definite models of the background, so that data might rule out a 
background interpretation.  But then the background model must be impeccable. This is 
why direct dark matter searches will publicly commit to their backgrounds and describe 
them in detail. When it comes to direct dark matter searches {\it the background IS the 
hypothesis} set up for the test. 

The literature shows that experimenters have been diligent in calibrating their 
instruments not to miss a signal. Discovering ``new physics'' always involves inverse 
questions that are more interesting than the calibrations.  The inverse question of 
neutron backgrounds is {\it given a putative signal not predicted by your neutron 
calibration, how will you know an un-calibrated kind of process did not make 
it?} This question is especially important for neutrons, whose interactions display great variety and rapid energy dependence. Yet the inverse question is remarkably undeveloped. There is unfinished work in the ``exhaustive characterization'' of backgrounds needed for experiments to make a convincing case. 

This paper contains provocative information of an interdisciplinary kind. It is due to contradictions between ``neutron physics'' cited in dark matter searches and neutron physics found in neutron-nucleus reactions. We might hear from one group or other that ``there are no neutrons, so don't worry.''  Plenty of literature show neutrons exist, and {\it experiments will eventually have to be seriously concerned}, whatever the degree of denial now. Our mission is to document misleading or incomplete information that has been used, gaps between what has been done and what is perceived, and consequences of not using information that is known.  

The author is not a member of any dark matter collaboration. Nobody but collaboration members know what the collaborations discuss among themselves. It is important this does not matter. Such things cannot make any difference in comparing {\it what was reported} with what exists. The {\it internally self-consistent} treatment of neutrons found in dark matter circles sometimes causes a certain complacency. For example the estimates of neutron fluxes from experiments - which range from ``no effect'' to ``the main background,'' depending on what is read - consistently refer to a special kind of neutron that was calibrated.  We will find those special neutrons do not really exist.  Then outside information about real neutrons often upsets the estimates. The contradictions put readers in the position where asking serious questions is appropriate. This does not take away from the dedication of experimenters to getting things correct. Asking questions of the experimenters is not supposed to need a license, and if some of our concerns turn out to be over-conservative, there is every reason to put them on the record in case they matter elsewhere. 

It is also not our burden of proof to show our proposals are final. The burden of proof 
goes the other way, and is held by the experiments making claims. We will show that 
``undiscussed problems'' cast serious doubt on claims that backgrounds have been 
understood. The only way dark matter can ever be discovered is by full disclosure 
and full participation of a community allowed to express criticism. 

To summarize the paper: We find the existing hypotheses about backgrounds stated by 
experiments, accepted at face value and as published, make possible a variety of 
neutron-induced events to be registered as dark matter signals. 

\subsubsection{Alternative Interpretations} 

The gaps in the literature about neutron-induced backgrounds suggest alternative 
interpretations. While anyone outside a collaboration tends to be limited to order of 
magnitude estimates (Section \ref{sec:props}), the alternatives are interesting in their 
own right, because they take into account new information:   

\begin{itemize} \item We propose that the rise of response below 1 KeV observed by 
CoGENT\cite{cogent2010} is the same rise seen in high-purity Germanium detectors 
from Auger-M electrons. Neutron capture and activation can produce it. 

Some dark matter literature will suggest that CoGENT has the finest energy resolution 
ever documented, making direct checks of our proposal difficult. However we find that 
much better detector resolution has long been routine in x-ray physics. In x-ray 
experiments a rise that appears to be CoGENT's signal has been observed (Fig. 
\ref{fig:acogpapover.eps}, Section \ref{sec:cogent}). If it is not the same signal, it {\it appears} to be the same signal, and we can't find any previous discussion.

To check our proposal we suggest a duplicate of the CoGENT experimental elements be 
set-up and exposed to a wide-band neutron and x-ray sources. It has not been done before. It will undoubtedly 
involve sacrificing some detector components to activation.  It is hard to know how
one would ever know for sure the signals were not the same without a direct experimental exposure. It seems remarkable that 
experiments whose entire value is determined by uncompromising background studies 
would not have planned sacrifical ``scorched-earth'' calibration from the start. 

\item We propose that events from the CDMS experiment are consistent with resonant 
neutron absorption and activation. CDMS reports that certain of its events cannot in 
principle be separated from neutrons, making the experiment dependent on background 
simulations. Scaling the backgrounds of neutrons in a more realistic (and experimentally 
more conservative) way than the extrapolations reported by CDMS appears to explain 
the signal. 

To check this proposal we suggest the full range of neutron interactions be incorporated 
in numerical simulations, which as we show below, has not been done to this date.

\item We propose that Auger electrons delivering about 3.1 KeV from activation and 
decay of Iodine-128 might explain the signal reported by DAMA/LIBRA. There is no 
record of previous consideration. The resonant features of neutron-induced backgrounds in $Na(Tl)I$ also present channels for direct excitation of the 2-6 KeV signal region. A seasonal variation of underground muon-induced 
neutrons is known to exist. We will show the time dependence reproduces DAMA's annual modulation with no free parameters. 

To check our proposal, we suggest that annual variation of $Na(Tl)I$ be duplicated in the Southern Hemisphere, perhaps at ICECUBE\cite{albrecht}. The experiment operates under thousands of water-equivalent meters of a low radioactivity, well understood substance, which is water. The backgrounds in ICECUBE are so radically different from Gran Sasso that discovery of seasonal variation in absolute phase agreement with the DAMA/LIBRA signal would likely be a {\it positive discovery} of dark matter not explainable by any known background. 

\end{itemize} 

\section{Three or More Misconceptions}  

Direct detection experiments operate in challenging low energy, low background regions 
never measured before. There is a feature of calibrating old physics and detecting new 
physics in one and the same experiment. In reading the literature three important 
misconceptions were found repeatedly. The origin is unknown, but the most cited 
reference is probably the 1985 paper of Goodman and Witten\cite{gooWitt}($GW$). 
The misconceptions are supported by textbooks, but textbooks about neutrons are 
sometimes wrong. 

{\it Misconception 1} amounts to transferring a model of dark matter signals to 
neutrons. Many years ago it was noticed that the elastic recoil of dark matter on nuclei 
could be imitated by elastic recoil of neutrons. Perhaps by subtle transfer of focus, 
we find an habitual transfer of the signal-model to the background-model has been used to effectively 
define neutrons. To be specific, $GW$ state a rule for the $2 \ra 2$ elastic scattering 
cross section of dark matter on nuclei\cite{gooWitt}: \ba \sigma_{GW} = {|M|^{2} 
\over (m_{1}+m_{2})^{2}  }, \label{sigGW} \ea where $M$ is the invariant scattering 
amplitude. A perception has come that all low energy cross sections take the form very 
generally, perhaps because it is so simple, but the formula is wrong in general. As a corollary 
comes {\it Misconception 2}, which is the energy losses of elastic $2\ra 2$ scattering 
computed for dark matter signals as well as neutron backgrounds by one uniform 
method. Given a target of mass $m_{2}$ at rest, struck by a particle with mass 
$m_{1}$ and Newtonian kinetic energy $E$, the energy transfer $\Delta E$ is predicted 
as \ba \Delta E  =2 E_{1} { m_{1}m_{2}  \over (m_{1} + m_{2})^{2} }   (1-cos 
\theta_{CM}). \label{scatt} \ea 
$GW$ actually cite the simpler upper limit\cite{gooWitt}. Then a 100-1000 GeV dark 
matter particle moving at speeds of order $10^{-3}c$ will produce recoil energy 
transfers at the 10-100 KeV scale. This is why direct dark matter detection has been 
driven into a challenging low energy region. Likewise, the formula predicts energy 
transfers to nuclei at the 10-100 KeV scale from MeV-energy neutrons in the lab by 
selecting $CM$-scattering angles $cos\theta_{CM} \ra 1$. This is very convenient for 
purposes of calibration, and used extensively. Indeed dozens of papers on ``quenching,'' which is the measure of elastic recoil energy, use Eq. \ref{scatt} as a definition of their study. While experimental conditions can be arranged in the lab to measure elastic scattering, it is a poor model for {\it all} neutron interactions that might contribute backgrounds.

Textbooks\cite{text} reiterate the $2\ra 2$ elastic scattering model for neutrons with 
statements that ``Elastic scattering from nuclei $n(A, n)A$...is the principal mechanism 
of energy loss in the MeV Region...In order for the inelastic process to occur, the 
neutron must, of course, have sufficient energy to excite the nucleus, usually on the 
order of 1 MeV or more. Below this energy threshold only elastic scattering may 
occur.''  
Detecting dark matter by arranging inelastic nuclear transitions was discussed by $GW$, 
Ellis {et al}\cite{Ellis}, as well as Engel and Vogel\cite{vogel}. Ref.\cite{Toivanen:2009zza} describes a shell-model calculation, and Ref. \cite{Xe129Inelastic} describes 
an experimental search. It is true that exciting a stable nuclear level will be inelastic. Yet the condition {\it the neutron must, of course, have sufficient energy to excite the nucleus} 
is wrong and defines {\it Misconception 3.} We'll see there is no lower limit on neutron 
energies to be inelastic, and much of what is roughly measured as ``elastic'' neutron scattering actually has an inelastic component depositing energy in the medium. The loophole exists in careful textbooks\cite{text} if buried under a qualifier ``...unless there are resonances.'' Nevertheless, a 
large body of work assumes neutron-induced inelastic events are taboo, and use a joint 
elastic billiard-ball framework where Eqs. \ref{sigGW}, \ref{scatt} are the starting 
points 
{\it of both signal and background estimates.}  

There is abundant evidence these issues affect dark matter searches:

\subsubsection{Calibrating On the Signal}  Every experiment calibrates dark matter 
signals using neutron scattering selected at the elastic point, and at a special neutron 
energy. None measure instrument response to the full range of neutron interactions 
\footnote{Collaborations hold the responsibility to disseminate the 
basic elements of their case. ``None measure'' refers to dozens of CDMS, CoGENT, XENON, DAMA/LBRA and 
dissertation 
studies, cited as the basis of claims for discovery, as well as the broadly cited trend of 
the literature, and its sub-literature. Communication with collaboration experts has been reported here to the extent that responses were received. }, and none put the detectors into situations to actually 
measure the full range of {\it background events} neutron interactions can produce. Experiments should state up front that the energy deposited by neutrons has not been measured and is unknown. This 
single fact proves our point that experiments have not made their case of exhaustively 
eliminating backgrounds. 

The neutron source $^{252}Cf$ has a spectrum sharply peaked around 1 MeV, and is 
used by every experimental group except CoGENT. Figure 3.9 of the 1996 dissertation 
of CDMS collaborator P. Barnes\cite{disserts} shows beautifully the famous 2-slope 
response in a 60 g CDMS prototype. The response is produced simply by recording 
events with $^{252}Cf$ (some neutrons, some gammas) compared to $^{241}Am$ (no 
neutrons, all gammas and alphas) while recording ionization and phonon energy. The 
ratio calibrated with MeV-scale neutrons remains a primary signal discriminator through 
generations of Ph.D. theses to this day\cite{disserts}. We can find no record of more 
general 
experiments, and even the use of another (MeV-scale) radioactive source is unusual. 
Note only is a single source, single calibration the rule, but even 
re-calibration with the a given source is done rarely, to avoid activating the detector. 

DAMA/LIBRA ($DL$) experiments use $^{252}Cf$ and also calibrate quenching using an MeV-scale 
neutron beam scattering at selected forward elastic scattering angles
\cite{chagani,damaapparat}. There are many such experiments, but always {\it 
selecting} elastic scattering at energies far above most dark matter backgrounds. 
DAMA's Ref. \cite{dama98} explains why generic neutron sources are avoided - to 
reduce activation. DAMA has never calibrated across the full range of neutron energies. 
``Pay 
attention that the aim of these calibrations (and others) is to induce nuclear recoils
with kinetic energy from few to several tenths keV in the detector, just to study the 
response of the detectors to recoils. Of course, to do that, neutrons of MeV energy are 
required\cite{bernabei}'' It is consistent with the doctrine that {\it neutrons only produce the recoil 
signals we use to calibrate.}

CoGENT avoids $^{252}Cf$ and calibrates on a monochromatic 24 $\pm 2$ KeV 
neutron beam using an ancient trick known as an ``iron filter.''  P. Barbeau
\cite{disserts} states that just three scattering angles were selected, to avoid damaging 
the crystal, and no other neutron exposure was arranged, to avoid activation. 

The XENON experiment\cite{manzurXenon,disserts} also uses $^{241}AmBe$, a source of MeV-scale neutrons. Its neutron beam calibrations use 2-4 MeV neutrons and time of flight to select elastic neutron scattering for calibration. Regions outside the elastic point are rejected. 

Once detectors have not been put into neutron beams of all energies to know the response, we cannot agree that the response to neutrons is known.

The exception to single-energy calibration was an experimental accident. It is the sole 
record {\it found published} among dark matter experiments actually exposed to 
moderated neutrons. The 2006 dissertation of CDMS-collaborator Michael Attisha
\cite{disserts} made the discovery with the SOudan LOw Background Counting Facility 
(SOLO). $SOLO$ was counting several blocks of low-radioactivity polyethylene for the 
CDMS experiment. A gross unruly response covering the whole detectable region of 
0-700 KeV was observed. The ``signals'' were traced to a photon cascade from a 
distant neutron calibration source moderated by the polyethylene bricks. Unfortunately 
the phonon signal from the moderated neutrons was not described. 

To repeat: The 
experiments on dark matter generally do not measure or report the response of the 
detectors to low energy neutrons. Response outside the region measured is based on 
modeling and theoretical extrapolations.
 
\subsubsection{What Happens at Resonances? } Dark matter experiments tend to adopt a 
kinematic framework, the ``rigid elastic billiard ball model,'' in one uniform way both for 
dark matter and to define known neutron backgrounds. On that basis neutrons are operationally  
``defined'' by the class of events found in ``quenching'' measurements. Anyone 
reading the experimental reports will find the elastic scattering formulas repeated again 
and again for backgrounds. It substitutes a kinematic model (Eq. \ref{scatt}) for a dynamical question.

Dynamics makes a difference. Even in a classical molecule made of balls and springs, 
the energy and momentum transferred depend on the time scale of interaction and the 
dynamics. For example, Eq. \ref{scatt} predicts a neutron energy loss $\Delta E_{n}  
\sim (m_{n}/m_{T}) E_{n}$ when the neutron mass $m_{n}$ is small compared to the 
target mass $m_{T}$. This predicts that $\Delta E_{T} / E_{n} \sim 1.3\% 
$ 
(for Germanium) and $\Delta E_{T} / E_{n} \sim 0.8\% $ (for Xenon). Conversely, if 
one believes the model, and observes an event with energy $\Delta E_{T}$, one is led 
to focus on neutrons from a particular narrow energy range, (or excluding them) declare a ``signal.''  Yet $\Delta E_{n}/E_{n}=1$ is allowed by conservation laws when a 
neutron hits a single nucleon inside a nucleus, which is very typical. (In pool, the cue ball (neutron) stops 
dead when elastically scattering on an equal mass ball (nucleon), while conserving 
kinetic energy and momentum. ) {\it After the event} the reaction of the struck nucleus 
as molecule will be whatever dynamics determines. It may re-emit the same neutron, or a different neutron, conserving total energy and momentum with an almost arbitrary amount of radiation. Such reactions may still be called ``elastic'' in neutron-physics parlance not overly concerned with radiation as a ``small effect.''

For comparison with rigid billiard-ball theory, Figure \ref{fig:GeSi(n,gam).eps} shows 
neutron capture cross sections in Germanium and Silicon. (When cross sections and 
decays are cited here, they come $nndc$\cite{nndc}, unless otherwise noted.) The 
wild 
energy dependence and sheer magnitude of cross sections defy the simplistic calibration 
procedures reported in the literature. (Note the log scale, with cross sections exceeding 
10,000 barns). 

\begin{figure}[htb]
\centering 
\includegraphics[width=3.5in,height=4in]{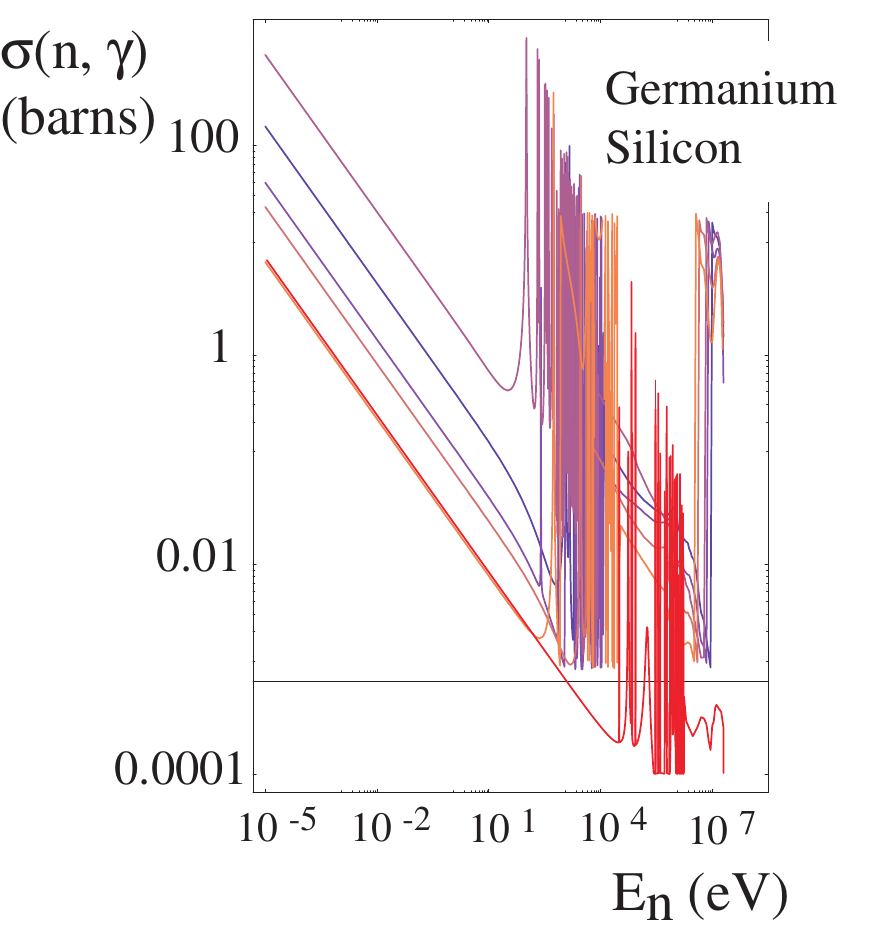}
\caption{\small  Neutron capture cross sections $A(n, \gamma)$ in barns on 
Germanium and Silicon (red online), which have been described as ``similar.'' 
Top to bottom (at $10^{-2}$ eV): $A$=73, 
70,72,74,76, 28. Neutron energies in eV. Data from $nndc$, Ref.\cite{nndc}. }
\label{fig:GeSi(n,gam).eps}
 
\end{figure}

\begin{figure}[htb]
\centering

\includegraphics[width=3.5in,height=4in]{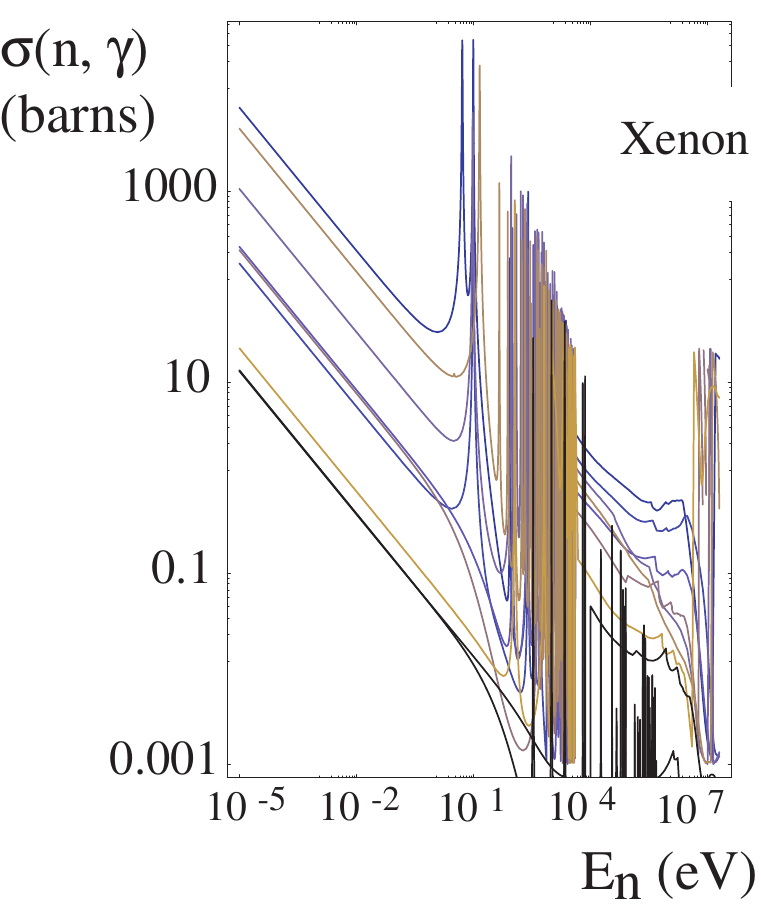}
\caption{\small  Neutron capture cross sections $A(n, \gamma)$ in barns on Xenon. 
Top to bottom (at $10^{-2}$ eV): $A$=124,131, 129, 128, 130,126,132, (134, 136, 
off 
scale at bottom). Neutron energies in eV. }
\label{fig:Xe(n,gam).eps}

\end{figure}

We naturally ask, {\it what is happening to the energy dissipated at all those 
resonances?} Each resonance has a width, which is the inverse time to decay. Energy 
is 
going into the medium; the energy in ``ionization'', or ``phonons,'' which are two 
names for quantum electrodynamic processes, will be spread around. The problem 
comes to why we can't find a mention. Every standard detector element, including 
Sodium, Iodine and Xenon have similar resonant behavior, in some cases even more 
dramatic (Fig. \ref{fig:Xe(n,gam).eps} shows Xenon). The thermal neutron capture 
cross section of $^{135}Xe$ (not shown, being produced by activation) exceeds 2.6 
{\it 
million} barns. This is 50 times larger than the famous 49,000 barn cross section of 
Gadolinium. The same effective cross section on $^{252}Cf$ (MeV neutrons) is only 
about 5.8 barns. Obviously the statistical fluctuations of events under such variable 
rates is a serious complication. The gross discrepancy between calibrating on one 
physics model and having backgrounds from different physics makes it fair to question 
whether backgrounds are known. 

Table 1 shows a small collection of numerical values of cross sections. It is not even 
clear which kind of cross section to use for an order of magnitude estimate. The 
nominal 
``elastic'' cross sections compiled for neutrons are\cite{nndc} (in general) the 
difference 
between the total cross section and the cross section to other channels that were 
measured. It is perfectly possible to distribute a few KeV of energy into the medium that was unobservable in old-fashioned neutron reaction experiments. We'll return to the question of ``what is meant by elastic?'' shortly. 

\subsubsection{What Compound Nucleus Theory Does Not Predict}  

Neutron physics has certain fundamentals. The theory called the ``compound nucleus'' 
explains the thicket of resonances. In inclusive neutron capture, denoted $A(n, \, 
\gamma)$, there is as much as 6-8 MeV of binding energy to dissipate, even when the 
neutron has zero kinetic energy. Normally one (or a few) MeV-scale gamma rays will 
account for most of the energy. Note that ``most of the energy'' is not ``all of the 
energy.'' 

Neutron physics would be much easier if dark matter experiments operated on MeV-scale signals. Notice that resonances start becoming important at neutron energies as low as 1 eV. 
There is no exact theory of the resonances, and they cannot be predicted on the basis 
of established low-lying nuclear excited states. The compound nucleus concept imagines
interaction with highly excited, densely packed levels 6-8 MeV above a bound state the 
neutron wants to make. This is strong interaction physics: it can be estimated in some 
statistical sense, but not predicted line by line. 

\begin{figure}
\centering
\includegraphics[width=3in,height=3in]{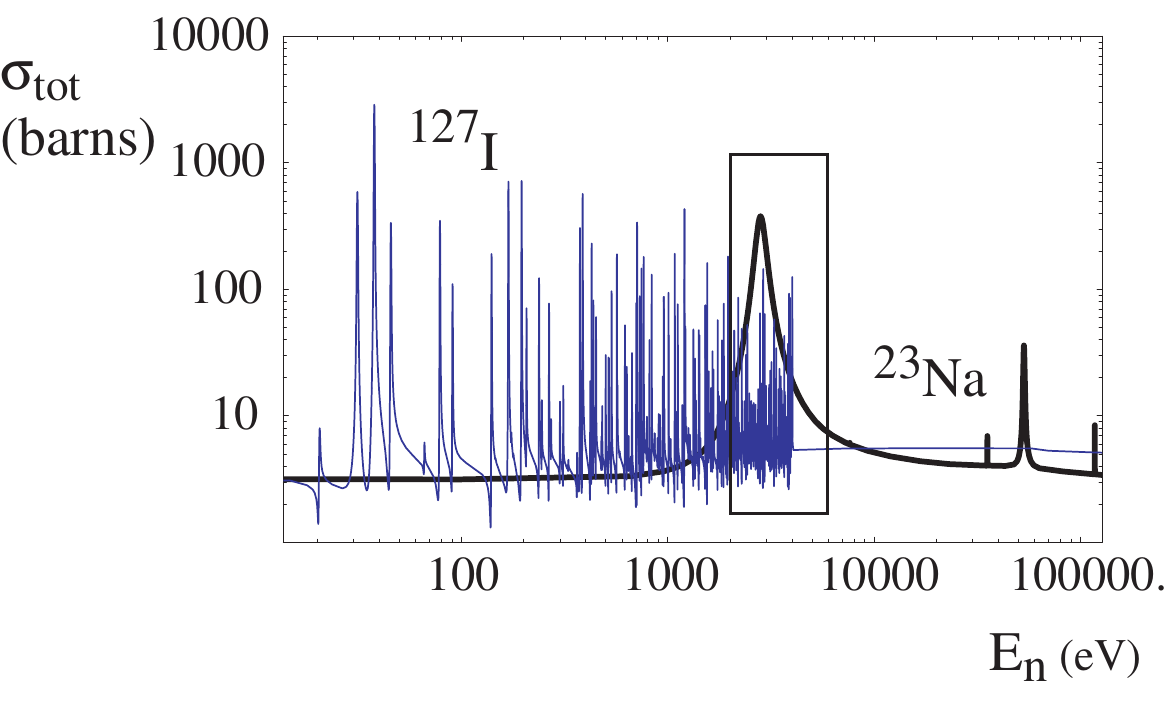}
\caption{ Resonant neutron total cross sections of neutrons in barns on Sodium (broad 
bump, black online) and Iodine (numerous spikes, blue online) versus neutron energy in 
eV. The DAMA/LIBRA signal region from 2-6 KeV is highlighted with a box.  } 

\label{fig:NaPlusICloseUpSigma.eps}
 
\end{figure}

Two different interpretations of the same words makes compound nucleus theory always correct by definition. According to tables the first excited state of $^{23}Na$ occurs at 439.990 KeV. Under interpretation-1 neutrons with energies less than 440 KeV (plus a bit for momentum) cannot possibly scatter except elastically. This kind of interpretation has been used in dark matter physics to ignore low energy neutrons. 
Yet data (Fig. \ref{fig:NaPlusICloseUpSigma.eps}) for neutrons on Sodium shows a whopping resonance at neutron energy close to 3 KeV. Interpretation-2 says the resonance comes when the neutron energy is close to an excited state of the ``compound nucleus''. Since the compound nucleus is defined by inspecting the data, the always-correct explanation is circular, and it means that whatever resonances are observed, are just the ones that occur. To repeat, it is completely wrong 
to think kinematics prevents a inelastic interaction below the energy to excite a stable 
nucleus. The correct rendition is that nuclear theory might hope to predict a resonance 
will occur when the theory of low-lying levels happens to apply. All the other resonances are a gift. 

Turn to atomic physics. A Bohr atom with $Z$ electrons has energy levels $E_{N} = 
m_{e}Z^{2}\a^{2}/2N^{2}= 34 \, KeV (Z/50)^{2}$. The inner shells are extremely 
responsive to KeV electromagnetic energy. The outer shells tend to act like $Z \sim 1$ 
and respond to eV-scale energies. Thus atomic energy scales overlap with the nuclear 
energy scales in complicated ways for the full range of resonances, and incoming 
neutron energies, $eV \lesssim E_{N} \lesssim 100 \, KeV$. The atom with $Z\a \sim 
1$ is a strongly interacting many body problem: we cannot compute what happens to 
the struck and resonating nucleus. 

Turn to data. In the range above 50 KeV (say) there are compilations of the photons 
coming out. For Germanium-73 there are 831 lines in the ENDSF compilation, and 415 
in the Budapest version. This is only one isotope of one element. We wonder why dark 
matter literature never mentioned it. The gammas complied from the single 9.47 eV 
resonance of $^{129}Xe(n, \, \gamma)^{130}Xe$ cover two pages of the journal
\cite{xenon947} in small font. However nuclear physicists take no responsibility for the 
low energy spectrum because it is not ``nuclear physics''. (To some extent {\it medical 
physics} compilations have proven useful!)  Not only are detectors inefficient below 100 
KeV - until the era of dark matter - but below 50 KeV photons tend to be absorbed 
inside materials more and more, until they don't come out at all !  The final answer to 
``what is happening at all those resonances'' to make up the total cross section is 
unknown. The cutoff seen in tables at low photon energies is not because gammas 
cease to be produced. {\it It is because detectors stopped measuring them. }   

Can energy losses be calculated from total cross sections?  In general, no. The 
combination of ordinary perturbation theory and a resonant cross section is often used 
in 
nuclear physics to estimate rates. It seems adequate for high energy (MeV) photons 
whose transition rates goes like a positive power of the energy depending on the 
multipole type. Energy losses from low energy radiation is not simple. On general 
grounds the ``total'' cross sections of processes with radiation is tricky. Low energy 
photon emissions are driven by the scale-free massless theory and tend to increase as 
energy decreases. The cross section {\it not to radiate} a massless photon actually 
vanishes due to infrared singularities - the Bloch-Nordseik phenomenon. Meanwhile the 
nominal elastic scattering cross sections of neutrons (as the difference of total and 
those with``hard gammas'') is actually inclusive. This is a significant point: the operational definition of ``nearly elastic'' neutron cross sections has always depended on the energy resolution. 

Experiments show that {\it The resonances of neutron scattering are accompanied by 
copious emissions of photons.} The total cross section tends to be a few 
orders of magnitude larger than the cross sections we've cited for capture. Whatever the complications of radiative damping that occurs, via resonance 
decay, internal conversion, bremsstrahlung, Auger electrons, x-rays, etc, it has been 
absorbed into the nominal elastic (i.e. inclusive) cross section found in tables, without 
being specific how energy is deposited. 

\subsubsection{Does the Energy Dependence of Neutrons Matter?}

To discover how neutrons affect experiments we look for evidence resonant processes 
were ever considered. Thousands of pages of Ph.D. dissertations\cite{disserts} and 
journal articles have been searched, and do not yield a single occurrence of the 
word ``resonance'' in the context of neutron-induced backgrounds. 

We also look for 
evidence that slightly different calibrations (despite the consistent agreement never to 
explore the full range) might have led to experimental inconsistencies. The dissertation 
of CDMS collaborator S. Kamat\cite{disserts} finds considerable inconsistencies 
between 
data and modeling that we can't find were resolved. The calibrations of quenching factors in Xenon by different groups using different methods do not agree: see Fig. 1 of Ref. \cite{xenon2010}, and also Fig. 8 of Ref.\cite{AprileXenon}.

We also discover that 
journal publications from dark matter experiments seldom if ever cite neutron cross sections, or do so only superficially. 
DAMA/
LIBRA cites a single thermal neutron cross section of 0.53 barns for a $23Na(n, 
\gamma)24Na$ triplle-coincidence used as  a neutron tracer. No CoGENT publication 
found cites a neutron cross section. CDMS\cite{cdms2003} writes: ``while Ge and 
Si have similar scattering rates per nucleon for neutrons, Ge is 5 - 7 times more 
efficient 
than Si for coherently scattering WIMPs.'' The dissertations\cite{disserts} of CDMS 
collaborators A. J. Reisetter,  R. Hennings-Yeomans, and C. N. Bailey, repeat the 
``similar scattering rates per nucleon for neutrons'' statement for Ge and Si almost 
verbatim. On the basis of repetition in dark matter literature the claim seems direct and well-documented fact.

Compare the claim with Table 1, which shows that Germanium and Silicon cross sections vary from 0.08 barn to 63 
barns, depending on how one chooses them. Fig. \ref{fig:GeSi(n,gam).eps} shows 
variations by factors of thousands. It is hard to see how Ge and Si are ``similar''. 
Comparing many other over-simple statements in dark matter papers with Figure 
\ref{fig:GeSi(n,gam).eps} and Table 1 suggest to us that dark matter collaborations are 
either not consulting the full range of information available about neutrons, or when 
they 
consult it, the facts become insider ``secrets.'' No discovery of dark matter is going to be based on secrets! Consider the consequences:  If there is more energy deposited than calibrated as neutron energies 
decrease, the current practice may count neutrons as dark matter. Conversely, if the neutron energy is overestimated by calibrations, than by miscounting neutrons the experiments may state false limits rejecting dark matter.  When there are two types of energy measured, such as ionization and phonon energy, then there are at least four (4) ways to go wrong with incomplete calibration.

\setlength\LTleft{0pt}
\setlength\LTright{0pt}

\begin{longtable*}{@{\extracolsep{1in}}p{4in}p{4in}}

 \begin{tabular}{cccccc}

\hline\\
 Nucleus  & \% & $\sigma_{thermal}(tot)$ & $\sigma_{ thermal}(n, \, \gamma)$  
(barns) &  $\bar  \sigma_{epithermal}(n, \, \gamma)$ (barns)  & $\sigma_{elastic}$
(barns)  \\ \\
   \hline \\ 
23Na& $\sim$ 100 &  3.84    &    0.53  &   	$\rightarrow $  0.32   &  1.6;  (2.45 
MeV)     
\\
28Si &  92.23 &   2.13   & 	0.16  & $\rightarrow $ 0.084   &  $ \sim 3\, ? \, $ 
(MeV)  \\
29Si & 4.67 &   2.70   &  0.12 & 0.08 & '' \\ 
30Si & 3.1 &  2.56   & 0.11 &   	  0.54  & '' \\ 
70Ge  & 21.23  & 16.9    &   3.10 &  2.60  &  $\sim  6\, ? \, $ (MeV) \\
72Ge  & 27.66 & 9.75    &  0.89 &  0.90   &    ''    \\
73Ge & 7.73  &   19.5  &  14.7  &   	$\rightarrow $  62.9    &          \\
74Ge  & 35.94  &  7.69   &  0.52 & 0.65   &       \\
76Ge   &  7.44  &  8.52    &   0.15   &  1.35  &        \\
124Xe & 0.09 &  150   &  150 &  	 3190 	 &        \\
126Xe	 & 0.089   & 8.53    &  3.46  & 66.5  &         \\   
128Xe	  &  1.90  &   12.0   &   5.19 & 12.66   &        \\
129Xe &  	26.4 & 24.6 &   21.0 & 245  &         \\
130Xe  &  	4.07    &  	11.0    & 4.78   & 4.91  &         \\
131Xe  &   21.2  &  91.2   & 90.0  &  $\rightarrow $ 882   &   ''     \\
132Xe  &   26.9   &    4.21   & 0.45  & 5.27  &   $\sim 2\, ? \, $ (2-4 MeV)      \\
134Xe  &   10.4  &  4.76   &  0.26  &  0.47  &    ''     \\
136Xe &   8.86  &    8.48  &  0.26  &   0.14  &    ''    \\  
127I &  $\sim$ 100 & 9.86   &   6.14 & $\rightarrow $ 159.6 & 3.2;  (2.45 MeV) \\  \\
     
\hline \\
 
\end{tabular}  

\\
 
\caption{A small survey of neutron cross sections. Natural abundance indicated by 
\%. The ``thermal neutron cross sections'' $\sigma_{thermal}$ are evaluated at the exact 
thermal energy 0.024 eV. The ``epithermal'' $\bar \sigma_{ epithermal } $ is the 
``resonance integral''; arrows show largest $abundance \times\bar 
\sigma_{ epithermal }$ for each element. Question marks indicates theory 
extrapolations at typical energies of elastic quenching experiments.
 }
  \label{tab:sigmaTable}

\end{longtable*}

\subsubsection{Reliance on Unverified Simulations} Lacking the data-versus data calibrations we 
first assumed was the core of backgrounds, we'd still hope numerical simulations took 
complete physics into account.

Unfortunately, numerical simulations do {\it not} incorporate even the fraction of 
neutron-inucleus-atomic processes that are known. Without being privileged to {\it all} 
simulation details, we can verify that {\it some} simulations are based on using high 
energy physics codes upon which the elastic $2\ra 2$ scattering of neutrons is tacked 
on 
at the end. Considerable attention has been given to generating neutrons from muons, 
which is a standard high energy physics task. For all the rest the coarse-graining of high energy codes 
customized for many-GeV events ought to be a concern. Propagation and interaction of 
neutrons through the resonance region ought to be a major complication, but it is 
another ``undiscussed problem.''

In the December 2009 paper\cite{CDMS2009} "Results from the Final Exposure...", 
CDMS states: " We performed Monte Carlo simulations of the muon-induced particle 
showers and sub-sequent neutron production with Geant4 [16, 17] and FLUKA [18, 
19].'' 
The code first calculates a neutron flux. The code then uses nuclear recoil theory, and 
calibrations on elastic neutrons, to classify events. It simulates some neutrons, not all 
neutrons. Actually the more detailed descriptions of CDMS simulations\cite{disserts} 
and 
CoGENT\cite{cogent2010} cite the neutron-transport code MCNP and MCNP-PoliMi as 
its 
benchmark of simulations. Pozzi\cite{Pozzi2003} describes them: ``In fact, in MCNP 
secondary photons are sampled
{\it independently from the type of neutron collision}.'' (Italics ours.) Confirmation that 
MCNP is not reliable for low energy gammas in $Na(Tl)I$ is given by Figs. 6, 7 of Ref. 
\cite{yazdi}. While there is a qualitative consistency, the calculations for thermal 
neutrons (the region not even including resonances !) do not look like the data below 1 
MeV. 

An improved code called MCNP-PoliMi\cite{Pozzi2003} exists. Unfortunately it 
faced a gap in dealing with a lack of experimental information on neutron scattering. Pozzi et al 
describe the resolution: `` ...The proposed
algorithm...searches for possible
cascades of up to three photons with energies
adding up to $E_{g}$...lack of
information forced us to select a procedure that
we considered reasonable. The advantage is that it
works in any case; the drawback is the introduction
of approximations that can be crude in some
cases.'' Here neutron processes have been {\it defined} to emit up to three (necessarily 
energetic) photons. Notice that ``up to three photons'' describes a procedure sampling low energy regions rather thinly - a  bin on the KeV scale is only about $10^{-4}$ of a 0-8 MeV energy range available. More reading confirms that MCNP-PoliMi does not claim to be 
reliable for MeV-scale photons. Using it for KeV-scale processes goes far beyond its 
design limitations. 

Perhaps groups are citing the public versions of MCNP and its variants
while they actually use their own (unreproducible) internal codes. CoGENT
\cite{cogent2010} describes MCNP-PoliMi used on their Gemanium detector without 
explanation. Hennings-Yeomans\cite{disserts} of CDMS describes extensive modeling 
with a modified MCNP-PoliMi. `` I have modified MCNPX (since
the source was made available) to include point-wise low-energy neutron scatters in
Ge and Si using the approach implemented in MCNP-Polimi [183], which includes
accurate simulation of the nuclear recoils. The MCNPX implementation was tested
using a monoenergetic neutron beam with an energy of 1 MeV .''  Note again the 
pattern to define neutron interactions by MeV-scale elastic scattering, and validate it self-consistently.

As for the core ingredients of MCNPX, its manual\cite{mcnpManual} reveals that the elements {\it Sodium, 
Germanium, Iodine, Thallium, and Xenon}, which constitute the detectors {\it all 
happen to be absent from the neutron cross section library.} Messages to the MCNPX physics subgroup seeking confirmation other versions don't exist received no response. 

We think this discovery gives reasons for concern. Experiments have wisely tried to 
minimize reliance on simulations. However, it has actually encouraged an impression 
that neutrons don't need to be modeled due to their inability to cause inelastic 
reactions.
The 1996 UC-Berkeley dissertation of CDMS collaborator P. Barnes\cite{disserts} writes that ``One 
line of defense against the muon-induced (underground) neutrons is to moderate the 
neutrons below detector threshold before they reach the detector. Note than an 18 
KeV 
neutron has a maximum energy deposition on germanium of 1 KeV. '' The same 
statements appear everywhere. If there is reason to {\it hope} that signals consistent with elastic recoil at MeV scales would still apply to multi-KeV neutrons, we don't understand 
how it is claimed to be {\it known}. 

\subsection{A Short List of Undiscussed Processes} 

The problem of energy losses is intimidating because Nature tends to find every channel possible. The Lindhard theory\cite{lindhard} describes the stopping power of atoms and ions in materials. It is based on non-relativistic particle interactions with a free-electron gas, assumptions that the interactions are a small perturbation, and applications of screened Coulomb collisions between two colliding atoms. Every dark-matter quenching experiment cites the Lindhard theory. Citation tends to be self-consistent because the experiments have been arranged to reproduce the theory's assumptions. 

None of the interesting neutron energy losses are included in the theory. Let us try to imagine a few alternative processes that may occur: \begin{itemize} \item Earlier we found the kinematics of neutrons and nuclei actually allows a neutron to stop dead, $\Delta E_{n}=E_{n}$, breaking all the rules of dark matter neutrons. \item After that, if the energy is ignored, the momentum transfer $\Delta p_{n} =p_{n}$ might appear in recoil kinetic energy, $\Delta E_{T} = p_{n}^{2}/2m_{T} \sim E_{n} m_{n}/m_{T}$, for which the elastic recoil estimate is ok. This depends on the time scale. \item However, if the time scale of the collision is very slow, the nucleus can be effectively pinned in the crystal lattice. A neutron at rest can fall in to become bound by negative energy $E_{B}$. The neutron accelerates to get momentum $\Delta p_{n} \sim \sqrt{ 2m_{n} E_{B}} \sim 10^{4}KeV/c$, and spreads this momentum among the other nucleons. The ``pinned'' nucleus absorbs $\Delta p_{n}$ and recoils with $\Delta E_{T} \sim  8 \, MeV/m_{T} \sim 80 $KeV. This calculation is completely different from the last, potentially yielding a signal. \item Nucleons typically have Fermi momentum of order 200 MeV simply from being bound. When a neutron arrives, it must fit itself into the wave function, by finding a way to distribute its momentum and energy in the face of Pauli blocking. However if the neutron is not captured, but merely scatters in a compound nucleus, it interacts (in theory) with excited states not subject to Pauli blocking. During that process the unstable system is radiating to develop the widths observed for resonances. \item A neutron with 1eV-10 MeV of energy might go into a nucleus and be directly captured. Since no detector is 100\% efficient, there is a finite probability not to detect the hard photon(s) radiated. The detector reacts to whatever energy it detects, in any form of recoil or resonance radiation, potentially making  a signal. \item In {\it electron capture} a nucleus with too many protons may also convert an atomic electron to a neutrino of energy in the MeV range. Let $E_{*}$ be the energy emitted with a massless particle. The nucleus recoils with energy \ba \Delta E_{T} =0.85 KeV \, \sqrt{ { m_{T}\over 72 \, GeV } { E_{*} \over 10 MeV} }. \nn \ea  The energy range below 1 KeV happens to be the ``signal region'' of the CoGENT experiment, and it produces a signal. \end{itemize} 

Do experiments with neutrons see anomalies not explained by the Lindhard theory? The work of Jones and Kraner\cite{joneskranerPRL,joneskranerOther} devised a rather ingenious method to measure atomic stopping power in Germanium. The experiments sought to detect recoil from MeV-scale photons emitted in neutron capture. A precise capture photon (915 KeV) was known to be associated with a decay pathway producing a particular tracer photon (68.5 KeV). The tracer photon was detected in coincidence with the Germanium crystal response. However foolproof the method might seem,  the results of $^{74}Ge$ compared to $^{72}Ge$ did not agree, essentially contradicting the Lindhard theory. (See Fig. 2 of Ref. \cite{joneskranerPRL}. Note that $^{74}Ge$ is de-emphasized on the plot. ) Some plausible excuses were constructed without really explaining the discrepancy. When the ``good case'' of $^{72}Ge$ was pursued in subsequent papers\cite{joneskranerOther} its energy deposition was found to be in discrepancy with theory by about 35\%. 

Note these discrepancies occur in experiments done with cleverly selected, precise lines predicted from first principles in order to make clean measurements. In using thermal neutrons they also avoid the dangerously resonant region entirely. It is unfortunate that information on the whole gamut of energy deposited in crystals of such experiments does not seem to be available.

\section{Our Proposals}    

\label{sec:props}
 
\subsection{DAMA/LIBRA} 

\label{sec:dama} 

DAMA/LIBRA is a 250 Kg $Na(Tl)I$ detector designed to exploit a strategy of 
overdetermination, internal-consistency and background rejection regardless of 
background origin\cite{damaEJP08,damaapparat,dama2010}.  ``Overdetermination'' 
means using multiple consistency conditions set up to be robust under revision of 
background estimates. DAMA does not use pulse-shape discrimination to separate 
recoils, and does minimal processing of raw data. Simulations and accounting for the 
overall rate and spectrum seen in the detector are sketchy. Kudryavtsev, Robinson, 
and  
Spooner\cite{kudrat2010} give a rather thorough overview of previous criticisms, stating 
that features of the reported spectrum are difficult to explain and that DAMA's divulging 
the measured energy spectrum up to MeV energies is necessary. We will also find that 
a lack of published information about the spectrum weakens DAMA's claims.

From DAMA's documents:\begin{enumerate}  \item The experiment seeks a time-
dependent signal. Annual variation of high statistical significance has been found. The 
phase of the signal is consistent with dark matter flux models. \item Accepted events 
must be detected in coincidence, by two independent phototubes connected to each 
crystal, which controls phototube noise. Multiple-hit events are excluded. 
 \item Environmental factors including temperature, electronics thresholds, air 
conditioning, and radon levels are continuously monitored with great care. 
Contamination with Radon, $^{238}U$, $^{232}Th$, $^{40}K$, etc. are 
acknowledged, with adequate documentation. \item The signal region in which annual 
variation is observed is limited to detected energy $2 \, keV< E_{d} <6 \, keV$. \item 
Any conventional background producing annual variation in the signal region should 
produce annual variation in the regions $E_{d}> 6\, KeV$, which is not observed. 

\end{enumerate} 

The crucial element, which is more critical and more ambitious than the 
procedures of other groups, is the concept that {\it any background producing a variation in the 
signal region should produce a variation over the full energy range above the signal 
region. } This is the extra structure to define a background hypothesis cited in the 
Introduction. The most important null regions of DAMA/LIBRA are the range $6 \, KeV 
< E_{d}< 20 \, KeV$, binned in Fig. 6 of Ref. \cite{dama2010}, and the measured rate 
integrated above 90 keV, denoted $R90$. 
 
The energy dependence of neutron reactions falsifies the crucial assumption. {\it It is 
perfectly consistent for resonant processes to affect one energy region more than 
others. }  

What are the important neutron-induced processes? DAMA/LIBRA cites 
a {\it thermal neutron} cross section of 0.53 barns for a $23Na(n, \gamma)24Na$ 
triplle-coincidence used as a neutron tracer. We agree with this particular cross section, 
listed in Table 1. Yet the rates for triple-coincidence are a strong function of detector 
acceptance, and also a tiny fraction of the total rate. Based on the number of 
coincidences detected, the group cites\cite{damaapparat,damaEJP08} ``An upper limit 
on the thermal neutron flux surviving the multicomponent DAMA/LIBRA shield has been 
derived as [110]: $<1.2 \times 10^{-7}cm^{-2} s^{-1} (90\% C.L.)$. The 
corresponding capture rate is: $<$ 0.022 captures/day/kg.''   For reference, the DAMA/
LIBRA signal modulation amplitude is about 0.02 $cpd/kg/KeV$, or 0.08 $cpd/kg $. A 
simple calculation confirms the flux and {\it thermal} cross section gives the rate of 
captures (counts) per day ($cpd$) cited. The context of DAMA's flux estimate is a 
consistency check designed to be an overestimate. For comparison Ref. 
\cite{wulandari} cites total fluxes (no shielding) in the Gran Sasso lab of order $1-4 
\times 10^{-7} /cm^{2}/s/MeV$ at the lowest (MeV) range measured. 

Total fluxes from rock are expected to be orders of magnitude larger than muon-
induced 
fluxes. But DAMA has a Lead brick shield inside the concrete, and Lead is an excellent 
source of neutrons. The production of neutrons from muons has been extensively 
studied by Kudryavtsev {\it et al}, including Ref. \cite{Kudy}. Mei and Hime's study of 
muon-induced backgrounds\cite{MeiHime} note that neutrons bounce around in caves, 
while stating ``We have seen a large increase (a factor 10-20, depending on the 
thickness
of lead) in the neutron flux due to the additional and
efficient production of neutrons in lead...''  Ref. \cite{nagai} actually observed 
$^{127}I(n, \, \gamma)$ in a 6.9 MeV line inadvertently induced in $Na(Tl)I$ by 
adding 
a Lead shield.

Why do groups emphasize thermal neutrons?  It is because they are simple.
Yet we have no reason to believe that muon-generated neutrons produced from the Lead or 
Copper 
(placed inside the concrete/polyethylene shielding) would be thermalized. Suppose they 
are {\it epithermal}: The resonance integral for Sodium is $\bar \sigma_{res}
(^{23}Na(n; \gamma) =$0.316 barns, and the estimate decreases. However the 
resonance integral\cite{nndc} for Iodine activation $\bar \sigma_{res}(^{127}I(n, \, 
\gamma) =$160 barns, about 300 times the figure used by DAMA/LIBRA. Iodine 
activation is not mentioned by DAMA/LIBRA other than a 1998 paper\cite{dama98} 
considering a triple-coincidence decay channel less distinctive than Sodium, and papers 
citing it\cite{bernabei}.

Investigation finds that the $^{128}I$ isotope produced by capture has a half-life of 24.99 minutes. The NNDC\cite{nndc} lists decays to 
$^{128}Xe$ (93.0 \%) and $^{128}Te$ (6.9\%) predominantly by emission of $
\beta^{-}$ and $\gamma$'s \cite{nndc}.  Beta-decays to $^{128}Xe$ are distributed 
with a mean energy of 801 KeV, and with $5.8 \times 10^{-4}\%$ in the region 
$67-242$ KeV. There is no surprise this would produce no detectable annual variation in 
$R90$, the integrated rate above 90 KeV. DAMA's claims that any events in this region 
would cause a variation are too broad to be well-defined and hinge on the parts of the 
spectra that apparently have not been published.

The decay to $^{128}Te$ gives definite reason for concern. Almost all decays (5.7 per 
100 $^{128}I$ nuclei) produce Auger-L electrons with average energy of 3.19 KeV. At 
the rate of 0.6\% the decay also makes x-rays with average energy 3.77 KeV, which 
will 
be immediately absorbed. These energies fit perfectly the DAMA/LIBRA signal region.

Decay to $^{128}Te$ also produces about 6\% rate to several Auger and x-ray lines 
over the region $22 \, KeV - 32 \, KeV$. This energy lies above the 20 KeV region of 
Fig. 6 of Ref. \cite{dama2010}, and so cannot affect it. The region is also absent from 
the region covered by $R90$. 

What can be done with the numbers?  Recalculating the background in terms of the 
0.022 $cpd/kg$ value with the resonance integral method gives \ba rate \, (\, 3 \, KeV \, )  \sim 
0.022 \, cpd/kg {160 barns \over 0.316 barns} \cdot 6\% \sim 0.7 cpd/kg. \nn \ea  
This 
is 10 times the DAMA/LIBRA signal. Assigning a 10\% annual variation gives a 
background well matched to the $DL$ signal. 

There is another way to make the estimate. Ref. \cite{damaapparat} states: 
``Assuming cautiously a 10\% modulation (of whatever origin)
of the thermal neutrons flux, the corresponding modulation amplitude {\it in the lowest 
energy region} (our italics) has been calculated by MonteCarlo program to be $<0.8 
\times 10^{-6}$ cpd/kg/keV ...the corresponding modulation amplitude is $<10^{-4} 
cpd$/kg/keV.''
 
Stating quantities in units of per-keV without controlling their energy dependence will not 
do. It is hard enough to estimate the total fluxes of neutrons. $DL$ presents a rough 
estimate of muon-induced neutron rates. It is based on 20 muons/$m^{2}/day$, 
times 
(1-7) $\times 10^{-4}$ neutrons/muon/(gm/cm$^{2})$, times 15 tons of material. 
({\it That} number of neutrons/muon comes from a measurement using liquid 
scintillator, a hydrocarbon with very low yield. ) $DL$ assume 1/16 of neutrons are 
detected, divided over 4 KeV, with a 2\% annual variation giving between ( 0.4 -3 )$
\times 10^{-5}$ n/day/kg/KeV. We redid the calculation with the following corrections: 
(1) 1.1 muons/$m^{2}/hour =26.4/m^{2}/day$ , the actual Gran Sasso flux; drop four 
factors of 1/2 used for detection of elastic scattering; multiply by 4 KeV artificially 
divided; use 0.3 neutrons/muon produced in 200 $gm/cm^{2}$ of Lead, cited for 
underground muons in Ref.\cite{Kudy}. The calculation gives \ba  annual \, variation 
&=&2\%  { 26.4 \, muons \over m^{2} day }{ 0.3 \, neutrons \over muon} effective 
\, 
area \nn \\ &=&  0.3 {neutrons \over day},  \nn \ea with an area of 1 $m^{2}$. If 6\%  of these neutrons give a 3 Kev signal the background from Lead-induced neutrons is 
0.19 cpd/250 kg= 7 $\times 10^{-5}$ cpd/kg. 

Perhaps these calculations are naive. They use numbers cited by DAMA/
LIBRA as straw-man overestimates made under the assumption that neutrons could not possibly matter. Similar complacency is observed in all elastic recoil-based estimates. 

Our case has been made when the uncertainties exceed the certainties. Once an order of magnitude calculation matters, a number of other factors enter.  DAMA has not published enough quantitative information to know whether the relative rate of multiple scattering to capture processes makes the veto of multiple-hits informative. What a reader can find is denial that neutrons matter based on the over-simplified picture of neutron interactions endemic to the dark matter community. 
Our estimate using epithermal fluxes is not particularly reliable, while capture cross sections themselves vary by thousands. If there is a significant energy deposition from resonant radiative damping 
perhaps the total cross section should be used. The {\it total} epithermal cross section of 320 barns rather than the capture cross section of 160 barns, doubling the estimate. Fig. \ref{fig:NaPlusICloseUpSigma.eps} 
shows that resonances become active in the DAMA/LIBRA signal region. Since it has not been published before, it represents work for the experimental groups using sodium iodide. What is known about how the 
the energy of those resonances is deposited? 

\subsubsection{Backgrounds Depend on Calculations}

Are the Auger calculations of 3.19 KeV energy at a 5.7\% rate reliable? Communications with the DAMA group indicate a disagreement about existence of the Auger L 
component. R. Bernabei responded\cite{bernabei} there is $\beta^{-}$ decay ``with 
end-point at 2 MeV. Again, no modulation has been observed at high energy, see for example the analysis of R90 in our papers. Therefore, the decay of 128I cannot play any role.''

We checked with experts\cite{sonzogni,firestoneprivate} that the 
1252 KeV $Q$ value of $^{128}I \ra ^{128}Te$ is carried away almost entirely by the neutrino. The small recoil energy of the nucleon inside the nucleus translates to the small Auger energy listed. The experts also explained\cite{firestoneprivate,sonzogni} that the Auger calculations 
given on $nndc$ should be good to relative orders of percent, while in the process of being made better
\cite{sonzogni}. The absence of the 3.1 KeV process on the LBNL website is because those process are not intended to be complete.\cite{firestoneprivate}.

Auger energies are first of all a weighted average of many 
complicated processes with their own internal nomenclature (Auger processes, for those not expert in atomic physics, are first labeled $K,\,  L, \, M $...from the inner shell-out.) They are then subdivided into multiple classes such as $ K-L2-L3$, etc. Different subdivisions and weightings cause small variations. Although sometimes obtained from measurements, the ``art'' of calculating Auger processes has been long established
\cite{sonzogni,firestoneprivate}. Much of the Auger ``data'' appearing in compilations 
has been replaced by theoretical calculations said to be highly reliable. 

When the DAMA collaboration was consulted\cite{bernabei}, we also confirmed that 
the detectors have never been calibrated with neutrons other than the few-MeV elastic 
recoil setups used for imitating dark matter signals. Indeed exposure to radioactive 
sources has been avoided to minimize activation. Ordinary $Na(Tl)I$ detectors have 
been exposed to neutrons and show a lively response, unfortunately not studied much 
below 1 MeV. While acknowledging that DAMA/LIBRA's multiple-detection strategy and 
background rejection have been set up to develop  a ``data versus data'' comparison, we do not find it convincing. Evidently what is known about the 
detector's response to low energy neutrons is based on extrapolations, Monte Carlo 
simulation, and rough estimates of Auger processes we hope might be improved in the 
future. 

On the basis of the information publicly available an activation process might as well 
explain the rate of DAMA/LIBRA's signals.  What about the time modulation?
 
\subsubsection{Seasonal Variation of Underground Neutrons}  

It seems not widely known that the ICARUS collaboration measured the seasonal variation of 
muons some years ago\cite{Icarus}. The data was recorded for about a year in Hall C of the Gran 
Sasso lab, using a 32 liter liquid scintillator detector. Events were classified as 
``alphas''
or "neutrons" with energies below (above) 3 (3.5) MeV. The authors of the study report an overall shift of the starting date by as much as 10 days may be possible\cite{raselli}, which is within the error bars. While this experiment is tentative and needs to be confirmed, the data available may be an important key to the DAMA puzzle.

\begin{figure}
\begin{center}
\includegraphics[width=3in]{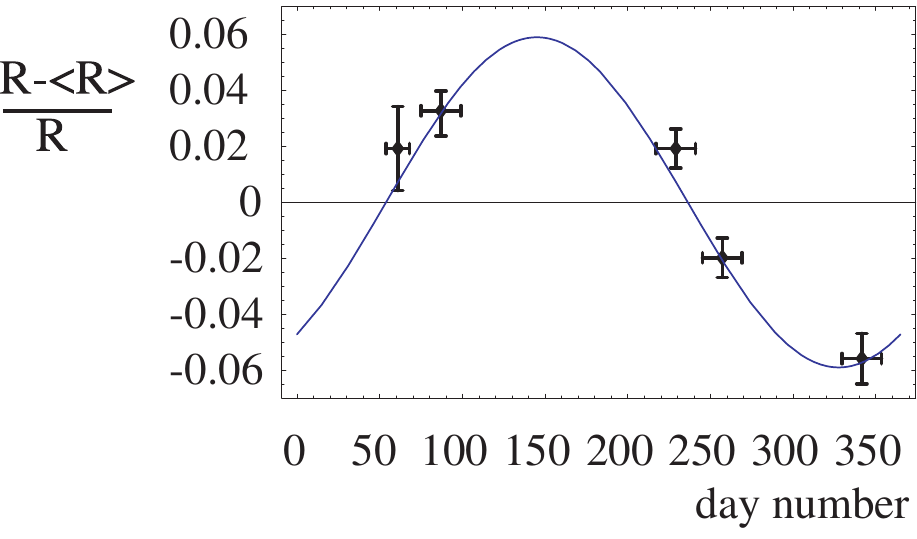}
\caption{ Time dependence of underground neutron rates $(R-<R>)/R$ in Hall C of 
Gran 
Sasso. Data from ICARUS collaboration\cite{Icarus} re-plotted on yearly scale. The cosine fit has its peak at day 146.4 $\pm 5$ , 
close to the peak of DAMA signals. }
\label{fig:IcarusErrorBars}
\end{center}
\end{figure}

Fig. \ref{fig:IcarusErrorBars} shows the ICARUS data for the rate deviation $(R-<R>) /R$ re-plotted on a cycle beginning with the calendar year. We fit the data to a constant plus cosine function, \ba {(R-<R>)  \over R }=   A cos( 2\pi (t-t_{0})/year). \nn \ea The fit gives 
$A= 0.059  \pm 0.01,$ and $t_{0} = 145.1 \pm 5$ days and $\chi^{2}/dof=1.1$. The peak of this data occurs on May 25, which is within one day of the peak observed previously for DAMA/LIBRA's 
annual signal modulation\cite{bernabei}. (More recently\cite{dama2010} the peak date 
cited is 152.5, which lies within the $\pm 7$ day error bars of previous estimates).

Encouraged by this, we propose that a ratio $f$ of the Gran Sasso neutrons are 
transmitted to the $DL$ detector in the range 1eV - 100 KeV. It is probably impossible 
to estimate $f$ from first principles. DAMA/LIBRA reports the apparatus is almost 
completely surrounded by 1 meter of concrete ``which acts as a further neutron 
moderator.''

\paragraph{About Moderation:} {\it Moderation} describes the process of slowing neutrons to 
room temperature by elastic scattering. Since the energy loss depends on the energy, 
it 
takes a surprising number of elastic collisions to bring neutrons to room temperature. 
Leo\cite{text} cites an analytic calculation for a 1 MeV neutron needing an average of 111 collisions with 
Carbon, and 17.5 with Hydrogen.  Neutrons actually evolve into an {\it epithermal} 
distribution phase\cite{WeinWig} in which their energy distribution goes like $1/E_{n}$ 
rather than being Maxwellian. As a result, the nuclear physics and nuclear power 
industry 
compile {\it resonance integrals}, which are cross sections integrated over the 
resonances and weighted by $1/E_{n}$. Comparing resonance integrals with 
``thermal'' 
cross sections is useful because it avoids the misconception that the thermal cross 
section (by definition evaluated at the single point $E_{n}=1/40$ eV exactly) might be 
a universal predictor - see Table 1.

Note that moderation does not imply absorption. On the contrary, {\it 
moderated neutrons are often assumed to be transmitted.} The ``moderating ratio'' of nuclear reactor theory is a weighted average of elastic to absorption cross sections. High moderating ratios are desirable for nuclear power uses, and indicate poor absorbers. Polyethylene has a very high moderating ratio\cite{rinard} of 122 in the 1 eV to 100 KeV range: it is a good neutron transmitter. Absorption is a different and specialized topic: there tend to be many channels, of which capture with $\gamma$ release ($A(n, 
\, \gamma)$) tends to be important. It turns out that Lead is among the top-10 best 
{\it transmitters} of thermal neutrons, with a tiny absorption cross section of 0.17 
barns. 
$DL$ also has a 1.5 mm layer of Cadmium between the concrete (outside layer) and 
Lead/Copper (inner layer). Elemental Cadmium has an absorption cross section of 2450 
barns, dominated by the cross section of $^{113}Cd$ (7.7\% abundance) exceeding 
20,000 barns, an isotope that happens to be radioactive. Above 1 eV the Cadmium 
cross section drops precipitously and its resonance integral is only 392 barns. Thus DAMA/LIBRA's envelope of 1.5mm of Cadmium is 
ineffective for epithermal neutrons. 

\begin{figure}
\begin{center}
\includegraphics[width=3.5in]{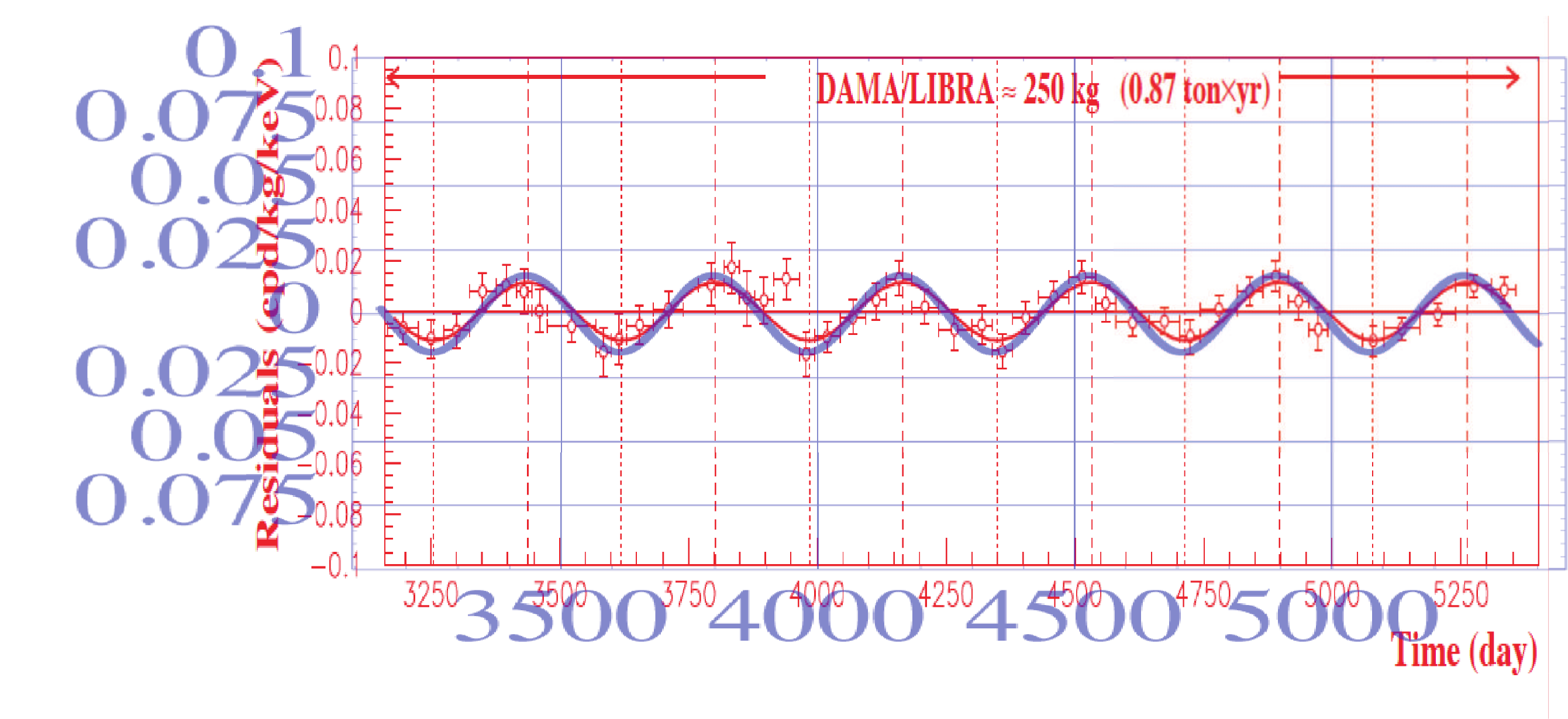}
\caption{ Overlay of time dependence predicted by the Gran Sasso underground neutron rate of Fig. 
\ref{fig:IcarusErrorBars} (blue online) with time dependence of DAMA/LIBRA signal region 2-6 KeV (red online, Fig. 1 of Ref. \cite{dama2010}). There are no free parameters in the period or phase. Overall normalization is free, and slightly misadjusted for graphic purposes. }
\label{fig:OverlapDama2-6KeV2010.eps}
\end{center}
\end{figure}

\paragraph{Continuing:} Now faced with the fraction $f$, we note that conservation of neutrons over the energy 
range $1 \pm 1$ MeV, where Gran Sasso neutron fluxes are measured, translates to an 
{\it increase} in epithermal flux at $10 \pm 10$ KeV by a factor of 100. Assuming this 
scaling and the known neutron normalization gives a prediction for the $DL$ single-hit 
residuals $A$, \ba  A = f \kappa \, cos( 2 \pi (t -145.1)/day).  \ea Here $\kappa$ is 
the conversion factor to obtain DAMA's definition of amplitude $A$. From the previous estimate the uncertainties justify fitting $\kappa$ to the data. Figure 
\ref{fig:OverlapDama2-6KeV2010.eps} shows the prediction from ICARUS time 
dependence for the DAMA/LIBRA signal in the range of 2-6 KeV. In order to retain 
information and avoid re-plotting errors, the prediction is overlain graphically with the 
original DAMA figure. Agreement is excellent. The normalization $f
\kappa=0.15$ was deliberately mistuned a bit to avoid a complete overlap.

Environmental neutrons from rock may not be the end of the story. It has long been 
known that underground muons show seasonal variation. MACRO\cite{macro} and MINOS\cite{minos} have measured variations at the few-percent level directly. 
MACRO's variation peaks at the beginning of July. These studies used selected ``high 
quality'' muons of a particular energy range, and are not represented to measure {\it 
all} muons. Higher quality selection produces smaller variation: the LVD collaboration
\cite{lvd} signal is only 1.5\%, but seems of very high quality. It peaks in July at day 
185 $\pm 15$. ICECUBE has reported variations between 10\%-15\% in ice at the 
South Pole\cite{icecubeSeasonal}. Once again this is high in South Pole summer, low in 
winter, exactly reversed in real time relative to the Northern Hemisphere. 

The history of annual variations is actually quite extensive. According to Rocco\cite{rocco} it was first observed by Forro\cite{forro} in 1947 before muons had been named. By 1952 Barret et al\cite{forro} gave calculations based on temperature variations in the 
atmosphere. The history of this effect indicates is not inherently small, while the 
amplitude as a function of energy can be debated, especially at low energy. It is difficult to assess the relative importance of rock-generated neutrons compared to 
muon-generated ones. A combination of the two might 
produce the signals reported by DAMA. Numerical work shows that the fit we cited is 
degenerate to shifting to earlier dates the peak of the rock-generated muons (as measured by ICARUS) while adding a compensating fraction of muon-generated neutrons peaking at 
day 185. This is because the sum of two cosines with different amplitude and phases 
is 
another cosine. Quite a large effect can be tolerated:  Letting ``1'' be the rock-
generated amplitude peaked at day 146.4, then (0.72, 0.41) are the amplitudes of 
rock-and muon components giving an equally good fit when shifted by 3 weeks. The whole range of fit parameters varying linearly with the shift will fit as well in between.  Thus DAMA's can be gotten by a wide range of annual oscillation effects and is not overly dependent on the ICARUS data. \footnote{Radon in Gran Sasso has significant annual variation 
documented by Bruno\cite{randonAnnual}. The peak at day 241 has no 
obvious correlation with the DAMA phase.}

\subsection{CoGENT}

\label{sec:cogent} 

Consider the CoGent experiment\cite{Cogent2008}. ``Listing from innermost to 
outermost components, the shielding around the detector was: (i) a low-background
$NaI[Tl]$ anti-Compton veto, (ii) 5 cm of low-background
lead, (iii) 15 cm of standard lead, (iv) 0.5 cm of borated
neutron absorber, (v) a $>99:9\%$ efficient muon veto,
(vi) 30 cm of polyethylene, and (vii) a low-efficiency
large-area external muon veto.''  

Note the 0.5 cm of borated neutron absorber. Below 10 KeV Boron is an effective neutron absorber, with the reaction $^{10}B(n, \, \alpha)$ dominating to total. At $10^{-2}$ eV (thermal) energies the capture cross section (weighted by abundance) is about 800 barns, making the 0.5 cm shield about 40 $ n_{23}$ interaction lengths, where $n_{23}$ is the number density in units of $10^{23}cm^{-3}$. However at 1 MeV neutron energy the cross section has dropped to only about 5 barns, making the 0.5 cm shield only about 0.25 interaction length. Fig. 9 of Ref. \cite{yazdi} is concerned with constructing Boron shields and has a discussion how Boron does not produce  complete shielding. We simply do not know the rate of multi-MeV neutron punch-through of the polyethylene and boron shield. Once again the Lead is on the inside, in part to shield from gammas produced by the neutron capture in Boron. Lead makes a neutron source for unshielded cosmic muons for which we have no background rate. No discussion of runs made with the crude but simple step of varying the thickness of moderator and absorber suggest an under-reporting (or immense confidence) about neutron backgrounds. 

CoGENT's signal\cite{cogent2010} is an unexplained rise in response below 1 KeV (Fig. 
\ref{fig:CogentPappComplete.eps}). Somewhat above the signal is a sharp line at about 
1.1 KeV attributed to $^{68}Ge$. The isotope is produced by knocking out alphas with 
rather high energy neutrons. The Barbeau 
dissertation\cite{disserts} states ``According to the background estimate, this would be the dominant 
contributor to the
background.''  That is partly on the basis of the isotope remaining in underground experiments with the somewhat long 271 day half-life after surface activation. The knock-out reaction has a cross section of about $10^{-2}$ barns at 10 MeV, (Fig. \ref{fig:72Ge(n,nalpha).pdf}), which then shrinks to negligible values below that energy.

\begin{figure}
\begin{center}
\includegraphics[width=3in]{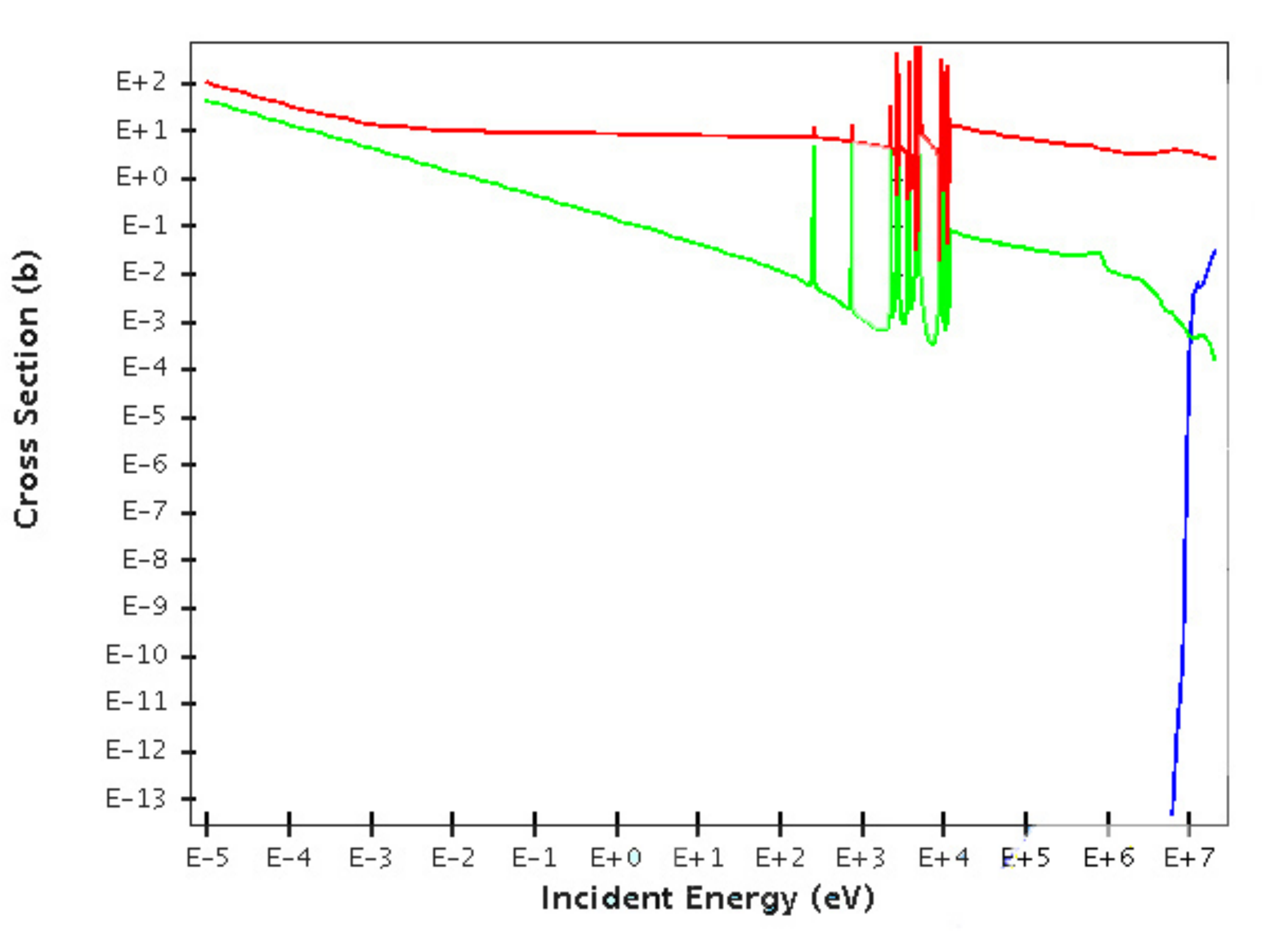}
\caption{ Cross sections in barns versus neutron energy in eV on Germanium: total 
(top), capture $^{72}Ge(n, \, \gamma)$, and $^{72}Ge(n, \, \alpha)$ (bottom right, blue online). 
Graphics from $nndc$ service $SIGMA$. }
\label{fig:72Ge(n,nalpha).pdf}
\end{center}
\end{figure}

We note that activation of $^{73}Ge$ and $^{75}Ge$ in the detector are not mentioned in any CoGENT document 
found\cite{barbeauJones}. Metastable $^{73}Ge$ is produced by activating $^{72}Ge$ (abundance 28\%, 
epithermal $\bar \sigma_{epithermal} \sim 130$ barns\footnote{The reaction $72Ge(n 
\gamma)$ has a standard resonance integral of 0.9 barns. The reasons for not 
saturating the reported total cross section are not clear. } and stable $^{73}Ge$ ($\bar 
\sigma_{epithermal} \sim 63$ barns. 

Metastable $^{73}Ge$ decays 198\% of the time(per neutron) to 1.19 KeV Auger-L 
electrons,  47\% of the time to 8.56 KeV Auger-K electrons, and there are about 50\% 
x-rays emitted\footnote{The ``percentage'' of photons emitted is per 100 days, and 
can 
add to a number greater than one.} at about 9.8 KeV. A nearly identical pattern of 
production and decay is seen for $75Ge$. The half-lives of metastable $^{73}Ge$ and 
$75Ge$ are 0.499 s, and 47.7 s, respectively. Since production and decay of these 
isotopes is too long delayed to veto with detector dead-times reported, the figures of high 
veto efficiencies do not translate into statements about neutron rejection. We then 
consider the activation processes as new candidates contributing to the background 
which contribute to in the line at 1.1 KeV. It is a reasonable proposal, given the low energy capture $^{72}Ge(n, \, \gamma)$ cross section is orders of magnitude larger than the cross section of the $^{72}Ge(n, \, n \alpha)$ reaction (Fig.\ref{fig:72Ge(n,nalpha).pdf}).

\begin{figure}
\centering 
\includegraphics[width=3in]{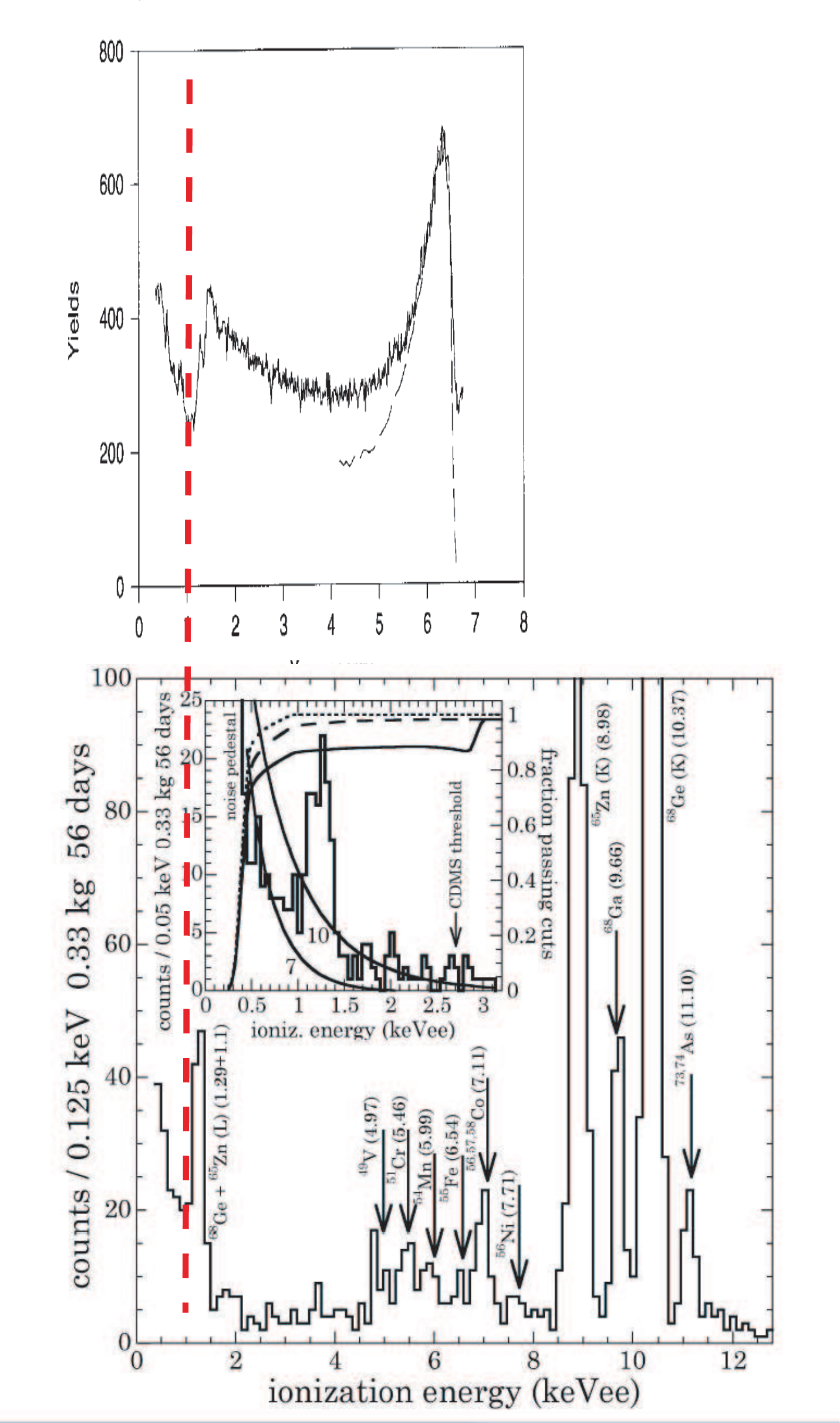}
\caption{ Comparison of the x-ray calibration of $HPGE$ detector from Papp 
2003\cite{papp}(top) to CoGENT 2010\cite{cogent2010} signal (bottom). The Papp figure has been 
stretched to the same horizontal and proportional vertical scale. Dashed line 
separates the CoGENT signal region, at left of line. In the same region Papp observes 
Auger-M photoelectrons. }
\label{fig:CogentPappComplete.eps} 
 
\end{figure}

Turn to the 8-9 KeV energy decay. The energy is so low it will seldom escape, but be 
converted inside the detector into other energy. We consulted the x-ray literature to 
find 
what happens. X-ray beams often have superb energy calibrations from coherent Bragg 
scattering, and they take data at tremendous signal to noise ratio. In fact Silicon and high-purity Germanium ($HPGE$) detectors have been resolving sub-
fractions of KeV energies in the x-ray regime for more than 20 years. Some theory of 
the response function is given in Ref. \cite{XRayTheory}. In 2003 Papp\cite{papp} 
published a detector study in which individual energy levels of atoms are beautifully 
resolved. {\it The stated detector resolution is 115 eV.}  Figure 
\ref{fig:CogentPappComplete.eps} (top) taken from that study is of great interest. 
Using 
an 8.4 KeV x-ray beam from CEA Saclay, the detector response reveals a low-energy 
ramp below 1 KeV. The paper reports this rise is {\it just as expected} from M-shell 
photoelectrons. The low-energy rise is not noise and is absent at lower x-ray energies 
(see discussion and compare Figs. 4a and 4b of Ref. \cite{papp}). Just to make sure this is real, Fig. 1 and 2 of Ref. 
\cite{lepy} show a similar rise. The x-ray response of the detectors is indistinguishable to 
our eyes from the CoGENT signal. 

\begin{figure}
\centering 
\includegraphics[width=3in,height=3in]{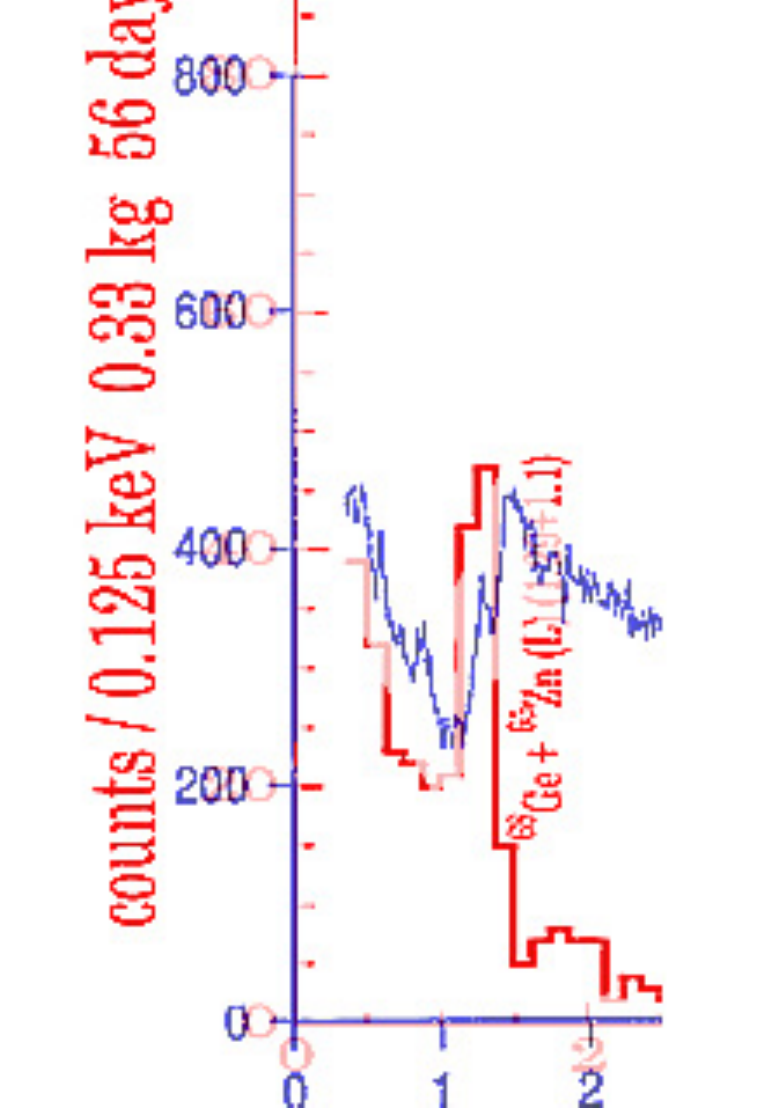}
\caption{ ``Signal region'' around 2 KeV selected from Fig. 
\ref{fig:CogentPappComplete.eps}. Graphics have been re-scaled and pasted one 
above 
the other to avoid re-fitting errors. CoGENT signal region is the rise below 1 KeV (red 
online). In the same region Papp observes Auger-M photoelectrons (blue online). }  
 
\label{fig:acogpapover.eps} 
 
\end{figure}

Why didn't CoGENT make the same proposal?  After we inquired, a collaborator\cite{barbeauprivate} expressed no concern for neutrons after ``shielding.''  An order of magnitude counting rule for Auger-M was cited. However the estimate and discussion do not appear in publications.  It is not consistent with the data of Fig. \ref{fig:CogentPappComplete.eps}.  Rather than quench the question, we find it is the burden of proof of groups such as CoGENT to show that this and other backgrounds have been considered and elminated.

Fig. \ref{fig:CogentPappComplete.eps} also shows that an 8.4 KeV x-ray beam excites 
some processes filling a region above 1 KeV not filled in the CoGENT data. Papp
\cite{papp} identifies the increasing component and peak around 6 KeV with the $KLL$ 
Auger electron spectrum produced by a Nickel electrode. The electrode is absent in CoGENT. There is also no reason to expect external x-ray irradiation to be restricted to the same effects as internal nuclear radiation, as we now explain. 

\paragraph{Problems of Internal Conversion:}  First, the CoGENT group is exploring a new technology of P-type point contact detectors 
for many applications, including double-beta decay. Their detector is designed to be 
more quiet than an ordinary $HPGE$ crystal detector by arranging very low capacitance, 
and corresponding low electronic noise (P. Barbeau, Ref\cite{disserts}). Since CoGENT's 
low-energy noise is small, it should be less than the noise of the first-quality $HPGE$ 
crystals of Ref. \cite{papp}, which already (to repeat) resolve the M-shell 
photoelectrons 
below 1 KeV. So both detectors see the same thing. Second, one of the main reasons 
that gamma-ray spectra from neutrons stop being tabulated below the 10-100 KeV 
region is {\it internal conversion}. The ``internal conversion coefficient'' ($ic$) of 
gamma rays is the ratio emitted to those absorbed by atomic electrons. Once gotten 
from data and hand-made calculations, $ic$ are now calculated on websites (``Hager-
Seltzer '' coefficients). At energies below 10 KeV calculations may be unstable. That is 
because $ic$ factors tend to explode as energies decrease: little to none of the energy 
escapes the atom. Clumsy efforts to generate and measure gamma-rays below 50 KeV 
escaping a substance may be an exercise in destroying the target. 

The coincidence of shape and strength of Auger-M lines makes us believe it may explain 
CoGENT's signal. There does not seem to be enough information published to verify 
whether rates from neutron-induced processes are consistent. Credible rate estimates 
will need a higher standard of simulation and a more complete reporting than CoGENT 
has published. Regarding neutron interactions inside the detector, T. Perrera
\cite{disserts} of CDMS makes a remark not seen elsewhere:``...most high-energy
neutrons will penetrate the polyethylene (moderator)...due to reflection by the lead 
shield, {\it these neutrons will pass though the
near-detector region many times before escaping, which greatly increases the 
interaction 
probability}.'' (Italics ours.)  Reflection seems plausible since Lead has little absorption 
and healthy (10-100 barn) estimated elastic cross sections. We believe if more were done or reported about neutrons inside the detector volume this remark would be resolved, but the undeveloped comment is the only case we have seen. 

Given the slim reporting and absence of direct calibrations with neutrons, we find every reason to 
propose that CoGENT has observed a neutron-induced background. 

\subsection{CDMS and XENON} 

Every experiment has an activation process we can't find discussed in the dark matter detection
literature. We find them by going to neutron physics data, and seeking big cross sections with decays in dangerous regions.

The December 2009 paper\cite{CDMS2009} "Results from the Final Exposure...", 
CDMS states that the cosmogenic background is estimated by a simulation. The 
observed number of vetoed single nuclear recoils in the data is multiplied by the ratio of unvetoed to vetoed events from the simulation. Note: {\it observed recoils are once again those events consistent with elastic recoil calibrations}... as opposed to all neutrons. 
Unvetoed events will be underestimated if they are simply absent from the code recognizing them.  

We then consider events from activation. Metastable $^{125}Xe$ is produced by 
activation of $^{124}Xe$, with cross sections $\bar \sigma_{thermal}$ =150 barns, $
\sigma_{epithermal}$ =3190 barns. It decays to itself (``isomeric transition'') with 
56.9 s half-life. Potential signals mimicking dark matter might be produced by Auger-L (3.43 
KeV, 88.7\%), Auger-K (24.6  KeV, 7.7\%), CE K( 76.7 KeV, 33\%) processes. 
$^{135}Xe$ is also neutron activated and decays on 15.3 minute half life. Once again 
the delay is too long for a muon veto to be effective. Apparently the most effective 
abundance-weighted cross section comes with $^{131}Xe$ (Table 1). It produces 
metastable $^{132}Xe$ with 8.4 ms half-life, decaying to 10-130 KeV Auger and x-ray 
processes about 50\% of the time. We are at a loss to explain why we've not found 
reports from the XENON experiment or associated dissertations about the resonant 
cross sections or backgrounds. As with the other information in this paper, we can do 
out best to read everything, but if a crucial basis of the experiment has been missed, it 
is really the collaboration's responsibility to publicize it. 

Earlier we mentioned that XENON calibrates its quenching factors with elastic scattering 
by neutron beams of MeV energies, selecting data from the elastic point. Fig. 7 of the 
paper by Manzur {\it et al} Ref. \cite{manzurXenon} is typical of the other studies. It 
shows that a slice of about 5 ns width was selected from a time of flight distribution 
over 90 ns wide. The cut selects the first half of particular bump in the data, under the 
assumption that the contribution from scatters other than single elastic ones is 
negligible, as supported by Monte Carlo simulations. Cutting to retain the data most purely suited to 
detect dark matter is done. Commitment to the elastic recoil model {\it for neutron 
backgrounds} is evident: ``In order to establish the background rejection
efficiency of LXe for a dark matter search, the absolute ionization and scintillation yields 
from nuclear recoils has to be precisely known\cite{AprileXenon}''. We commented earlier that different quenching experiments of Xenon do not agree. It is probably significant that the points of largest quenching and largest disagreement\cite{chapel} were taken at higher energies of 6-8 MeV. Comparing Fig 3 (top) of Ref. \cite{chapel} with Fig. 7 (bottom) of Ref. \cite{manzurXenon} finds entirely different structures identified as ``elastic scattering''.  If the different measurements can be reconciled, 
the consistent focus on elastic recoil is complemented by a possible lack of curiosity in  
actually finding out what the more general interactions of neutrons might be.

We propose there is work to do in calibrating Xenon with every possible background. 
Paying the price of activating a Xenon sample ought to help experiments calibrate the 
interactions of neutrons in order to support the case backgrounds are understood. For 
one thing, measurements would support simulations of the XENON active shield, which 
due to the complicated resonant interactions in conjunction with multiple scattering is 
not a simple affair. The ``First Dark Matter Results'' of XENON100\cite{xenon2010} report little more than
"GEANT4 Monte Carlo simulation of the entire detector", while GEANT4 is high energy physics code not designed to emulate the low energy interactions and propagation of neutrons. 

If what has been published is true and complete, so that the response to neutrons is based on extrapolations rather than direct calibration, then
it may also happen that events misidentified as neutrons and rejected could be due to dark matter. 

\subsubsection{ Is More Neutron Data Needed? }

The body of knowledge of neutrons experiments themselves give relatively little 
information about the resonant states of neutron scattering. It is not clear whether new experiments 
are needed. 

The {\it Atlas of Neutron Resonances}\cite{atlas} compiles thousands of experimentally 
determined 
resonance parameters collected over the past 75 years. Nevertheless, cross section 
data may not be sufficiently complete or reliable. 

\begin{figure}
\begin{center}
\includegraphics[width=3in]{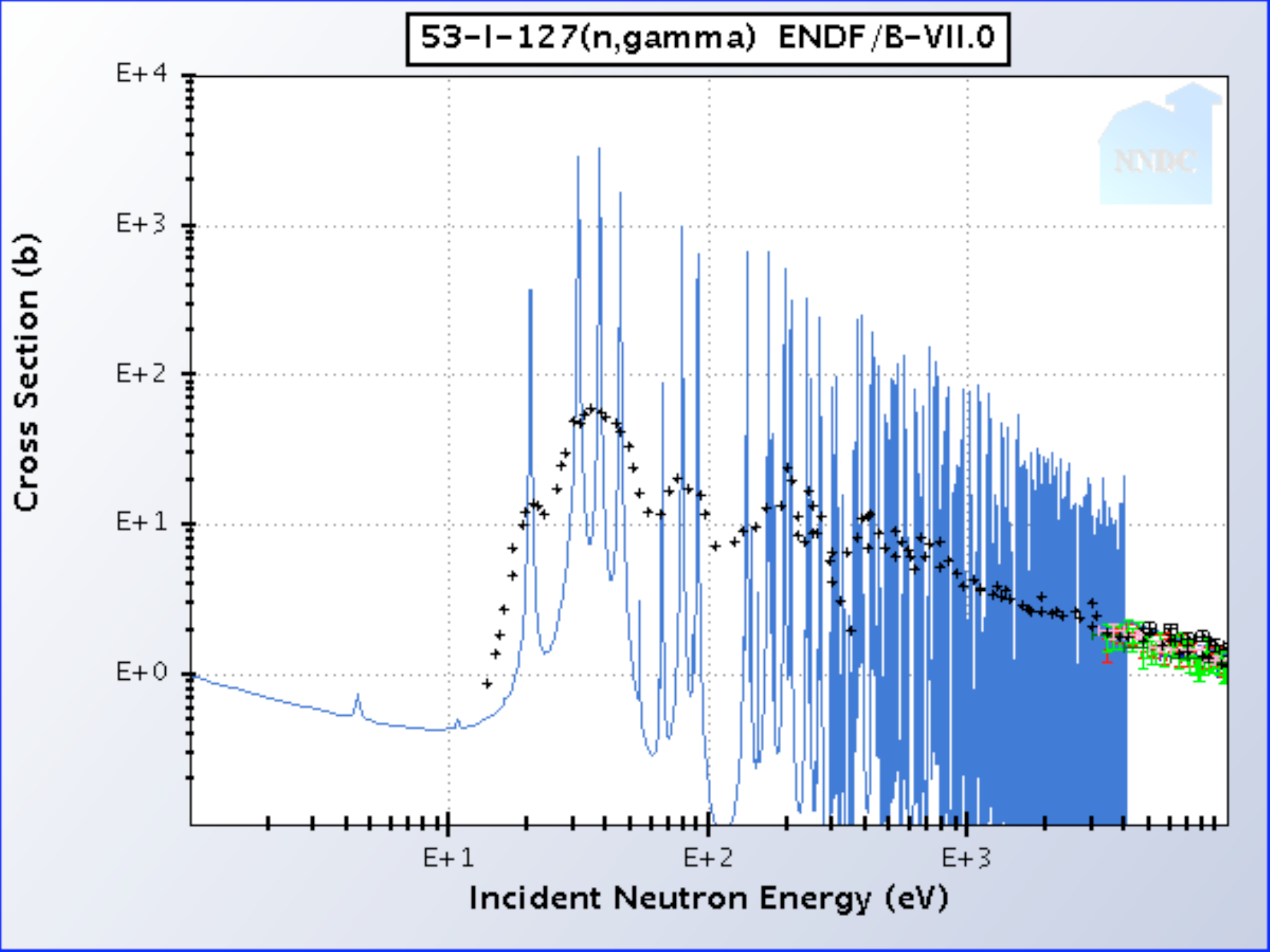}
\caption{Cross section data (points) and resonance parameter data of neutron capture 
cross section $^{127I}(n,\gamma)$ plotted by $nndc$. Different methods were used 
to measure resonances and cross sections. }
\label{fig:127INNDC.pdf}
\end{center}
\end{figure}

Much of the data is old. According to the author of Ref.\cite{mugab}, many 
experiments were done by transmission methods. They did not actually report cross 
sections, but fit data directly to multiple Breit-Wigner parameters, which were more 
convenient to report. Thus one will find gaps and possible discrepancies comparing 
actual cross sections reported, versus the detail shown by resonance parameters. And 
then, a cross section measurement does not necessarily disagree with a resonance 
parameter measurement, because different things were measured....Fig.\ref{fig:127INNDC.pdf} shows $127I(n,\gamma)$ cross section data and evaluated data on 
the a finer scale. (On the $nndc$ website $SIGMA$, one clicks ``Plot experimental data (EXFOR)'' and 
unclicks ``ENDF/B-VII.0 Library'' to get rid of the resonance parameters. In some 
cases 
this leaves a blank plot !)  We are assured\cite{mugab} by the author of the {\it Atlas} 
that resonance parameters ``are not based on any theory, but all come from 
experiment.''  

As mentioned earlier, neutron {\it elastic} scattering seems to have relatively scant data. It is sometimes assigned by models. The 
optical theorem relates the total cross section to the forward {\it elastic} scattering 
amplitude: \ba Im(M(s, \, t=0)) =4 \pi k \sigma_{tot}(s). \nn \ea As a consequence, 
rapid variations in {\it total} cross sections from thresholds and resonances also tend to 
be reflected in rapid variations in {\it elastic} cross sections. Total and elastic cross 
sections don't predict each other, but models exist; they are used when needed on the 
website.

While ``based on experiments'', we find the actual {\it data} on neutron cross sections 
and energy losses (not discussed enough in direct dark matter detection) may not be 
sufficiently reliable for background calculations. Recalling the claims that Ge and Si cross 
sections are ``similar'', we reviewed the elastic data. The actual set of {\it data} for 
Germanium elastic neutron scattering at $nndc$ consists of one (1) experimental point 
at 0.02 eV neutron energy that was reported at a 1970 Helsinki conference.
\cite{helsinki}. But at least the theoretical extrapolations for Ge and Si to the special 
kinematic calibration points of neutron quenching studies are ``similar.

Error bars on the energy deposition are not easy to find out. We reiterate the historical 
reasons for the lack of detail at low energies. Much of the data is quite old. The old 
technology focused early on energies of 1 MeV and above where the detectors were 
efficient. MeV-scale detection complemented old nuclear theories focused on the 
excited 
levels of stable nuclei. The needs of nuclear power and weapons research focused 
early 
on the net flow of energy, as dominated by sufficiently energetic processes. 
Resonances extending through the KeV region and below were historically hard to 
measure, and lacked a predictive theory. The ``Hauser-Feshbach'' statistical model 
came to be accepted as  ``good enough'' for most nuclear physics, and remains state-
of-the-art. Information on more sophisticated neutron energy losses is remarkably hard 
to find, and possibly classified secrets. This is underscored by the segregation of 
neutron experts and experiments at weapons labs. 

The information found falls short of the incredibly refined ambitions of dark 
matter experiments to know all background energy below 50 KeV and seeking cross sections below $10^{-44} cm^{2}$. 

\section{Summary} 

A review of the literature finds a variety of neutron background processes have been under-recognized or perhaps under-documented in the reports of major experimental groups. Since the premises of the experiments rely on quantifying backgrounds above and beyond theoretical extrapolations, neutron backgrounds cannot be dismissed.

It is sometimes thought that direct detection of dark matter is so exciting that the drive for discovery is guaranteed to dominate. However the hypothesis of particle dark matter is defined very loosely in its magnitude and rate. No experiment can falsify the hypothesis. The proposal can only be ``truthified'' by a dramatic contradiction to conventional physics. But if truthification had top priority, the procedures of DAMA/LIBRA might have been repeated already in the Southern Hemisphere. This indicates there are other forces at work. Nobody wants the field of dark matter detection to follow the slippery slope of unprogressive conservatism where higher and higher status will be achieved by developing better and better technology to discover nothing. 

In seeking one consistent explanation of everything known, it came as a complete surprise that the backgrounds, not the signals, are the actual hypotheses of the dark matter detection experiments. The experimentalists have to prove their case for backgrounds, not for signals. That is why discussion is appropriate. The situation is rather different from discovering new physics at the LHC (for example), where certain signals of new physics might stand out. In contrast, the dark matter experiments have already granted that nothing stands out in direct detection and nothing matters until all backgrounds have been exhaustively characterized. 

This is why backgrounds have become everyone's business, and sitting back to accept casual assurance can only harm progress. There are many cases in physics history where discoveries were late because backgrounds of conventional physics became the domain of technical specialists. For example, Leverrier discovered the precession of the perihelion of Mercury in 1859, contrary to rumors that Einstein predicted it in 1915. Leverrier's proposals failed, and the topic was left to backgrounds of classical celestial mechanics. More and more terms in perturbation theory kept pace with more precise measurements, and for 50 years the inherent profit of validating the status quo, fiddling with parameters, and never taking risks, blindly confirmed Newton's Laws to higher and higher accuracy. This might have continued forever, so that the opportunity to question the law of gravity itself might have never been born.

Not every experimental group queried showed a willingness to discuss anything about backgrounds. This is something to ponder. We showed here that the existing literature contains many lapses in the treatment of neutrons. The phenomena observed so far appear to be consistent with backgrounds amended to include known neutron physics. We think this is progress. 

It would be also progress for experiments to begin clarifying and publishing their neutron backgrounds as a matter of course. For many years collider experiments have published and released their continuum backgrounds, which are fit by thriving communities of higher-order QCD and numerical simulations. In both collider physics and dark matter, the backgrounds are the bulk of the data measured, are interesting in their own right, and cannot possibly be kept as proprietary secrets. Engaging everyone with complete transparency and openness will be how new physics might come to be discovered.

{\it Acknowledgments:}  Many experts generously gave information, including Richard Firestone, Alejandro Sonzogni, S. F. Mughabghab, Boris Pritychenko, G. A. Miller, Blayne Heckel, Craig Roberts, Jim Vary and J. D. Bowman. Thanks to Phil Barbeau, and Durdana Balakishiyeva for information they were able to provide. Rita Bernabei was extremely forthcoming and refreshingly straightforward. Yudi Santoso, Mihailo Backovic, and Danny Marfatia pointed out references and helpfully obtained copies. A preliminary version of this work was presented at Pheno2010, for which the organizers are thanked.  Research supported in part under DOE Grant Number DE-FG02-04ER14308.


\begin{thebibliography}{99}
\bibitem{albrecht} Discussions with Albrect Karle find the ICECUBE 
collaboration has this possibility under consideration.
\end{thebibliography}
\end{document}